\documentclass[man]{apa6}

\usepackage[american]{babel}
\usepackage{csquotes}
\usepackage[style=apa,sortcites=true,sorting=nyt]{biblatex}
\usepackage{graphicx,epsfig}
\usepackage{amsmath}
\usepackage{amssymb}
\usepackage{color}
\usepackage[makeroom]{cancel}
\usepackage{verbatim}
\usepackage{bm}
\usepackage{fancybox}
\usepackage{color}
\usepackage{mdframed}
\usepackage{tikz}
\usepackage{caption}
\usepackage{lipsum} % two abstracts 
\usepackage{threeparttable}
\usepackage{listings} %code
\usetikzlibrary{shapes.geometric, arrows} %flowchart
\usepackage{multirow}
\usepackage{threeparttable}
\usepackage{subcaption} %footnote for firgures
\usepackage{hyperref}
\hypersetup{
    colorlinks=true,    
    linkcolor=blue,    
    citecolor=blue,    
    filecolor=blue,    
    urlcolor=blue      
}
\DeclareLanguageMapping{american}{american-apa}
\title{Evaluation of Missing Data Analytical Techniques in Longitudinal Research: Traditional and Machine Learning Approaches}
\shorttitle{Missing Data Analytical Techniques}
\author{Dandan Tang and Xin Tong
}
\affiliation{Department of Psychology, University of Virginia
}
\authornote{This research was supported by the National Science Foundation under grant no. SES-1951038. Correspondence should be addressed to Dandan Tang. Email: gux8df@virginia.edu.}

\abstract{
Missing Not at Random (MNAR) and nonnormal data are challenging to handle. Traditional missing data analytical techniques such as full information maximum likelihood estimation (FIML) may fail with nonnormal data as they are built on normal distribution assumptions. Two-Stage Robust Estimation (TSRE) does manage nonnormal data, but both FIML and TSRE are less explored in longitudinal studies under MNAR conditions with nonnormal distributions. Unlike traditional statistical approaches, machine learning approaches do not require distributional assumptions about the data.  More importantly, they have shown promise for MNAR data; however, their application in longitudinal studies, addressing both Missing at Random (MAR) and MNAR scenarios, is also underexplored. This study utilizes Monte Carlo simulations to assess and compare the effectiveness of six analytical techniques for missing data within the growth curve modeling framework. These techniques include traditional approaches like FIML and TSRE, machine learning approaches by single imputation (K-Nearest Neighbors and missForest), and machine learning approaches by multiple imputation (micecart and miceForest). We investigate the influence of sample size, missing data rate, missing data mechanism, and data distribution on the accuracy and efficiency of model estimation. Our findings indicate that FIML is most effective for MNAR data among the tested approaches. TSRE excels in handling MAR data, while missForest is only advantageous in limited conditions with a combination of very skewed distributions, very large sample sizes (e.g., n $\geq$ 1000), and low missing data rates.

} 
\begin{document} 

\maketitle

\section{Introduction}
Longitudinal research plays a crucial role in social and behavioral sciences, offering a unique perspective by tracking the same subjects over time. This approach allows for analyzing changes within individuals and variations across different individuals in change. Such research provides invaluable insights into developmental and progressive phenomena, covering various topics such as children's emotion regulation (Zhu et al., 2023), cyberbullying perpetration (Camerini et al., 2020), sleep health (Deering et al., 2020), etc. Owing to its significant usefulness, the popularity and application of longitudinal studies have increased over the years. This increasing trend is evident in the growing volume of related studies appearing in databases. For instance, searches in the PsycArticles database for terms like 'longitudinal studies' or 'longitudinal research' reveal a steady rise in publications: 8,277 articles from 1994-2003, 14,591 from 2004-2013, and 26,061 from 2014-2023.

Despite the popularity of longitudinal studies, researchers often need to deal with missing data issues in these studies, mainly because data are collected over time, and participants are not always available to be measured at all measurement occasions. For instance, a meta-analysis of 92 longitudinal studies on personality traits found an average missing data rate of 44\% across the sampled studies (Roberts, Walton, \& Viechtbauer, 2006). Such incomplete data in longitudinal studies can lead to inefficient or biased parameter estimates (Ibrahim \& Molenberghs, 2009). For example, in a growth curve analysis, testing hypotheses using the Wald statistic with incomplete data could result in a type I error rate lower than the nominal threshold of 0.05 (Leeper \& Woolson, 1982).

To better understand the data missingness, Little and Rubin (1976) classified three types of missing data mechanisms: Missing Completely at Random (MCAR), where missingness is not related to any observed or unobserved data; Missing at Random (MAR), where missingness is related to observed data but not to the missing value itself; and Missing Not at Random (MNAR), where missingness is related to unobserved data. In the longitudinal setting, consider the relationship between sleep quality and academic performance among college students as an example. If, at one assessment wave, some responses from participants fail to record due to a glitch in the survey software, the missingness is unrelated to any participant characteristics or study variables and thus is considered MCAR. If the likelihood of missing data is related to an observed variable (e.g., course load), which is incorporated in the longitudinal model but not related to the unobserved variables of interest (i.e., sleep quality or academic performance), the missingness is MAR. If the students experiencing poor sleep quality are less likely to attend follow-up assessments due to feelings of fatigue or disengagement, the probability of missingness is related to an unobserved variable (i.e., poor sleep quality) that is directly linked to the variables under study (i.e., academic performance), classifying it as MNAR.  Since MCAR and MAR data typically do not require modeling the missing data mechanism to make valid inferences, they are also called ignorable missing data. In contrast, for MNAR data, as the missing data mechanism needs to be modeled directly to correct for the bias it introduces, it is called nonignorable missing data.

Various statistical techniques have been developed to tackle the issue of missing data. Convenient approaches like listwise or pairwise deletion may apply to MCAR data, but they often lead to a loss in statistical power due to reduced sample size and biased results for MAR and MNAR data. For longitudinal models in the structural equation modeling (SEM) framework, the Full Information Maximum Likelihood (FIML) is widely recognized. FIML is favored for its ability to provide consistent and efficient parameter estimates for MCAR and MAR data with normal distributions (Enders \& Bandalos, 2001; Shin et al., 2017).  However, practical data are often nonnormally distributed in social and behavioral sciences (Micceri, 1989).   In such instances, FIML may generate  biased parameter estimates and be inefficient (Yuan \& Bentler, 2001). The Two-Stage Robust Estimation (TSRE), proposed by Yuan and Zhang (2012), can be flexibly applied to nonnormal data by downweighting potential outliers (e.g., Cruz et al., 2023). Further research by Yuan, Tong, and Zhang (2015) demonstrated that TSRE could produce less biased and more efficient parameter estimates in cross-sectional, nonnormal data modeling with both MAR and MNAR mechanisms. Despite these developments, the application of FIML and TSRE in longitudinal data under MNAR conditions, particularly with nonnormal distributions, remains underexplored (Shin et al., 2016). Dealing with MNAR data remains one of the most challenging aspects of missing data analysis, as determining the underlying missingness mechanism without additional information is difficult, and many statistical techniques cannot effectively address this type of missingness.

Machine learning approaches, such as tree-based algorithms and the K-Nearest Neighbors algorithm (KNN), have recently become popular tools for imputing missing values. These techniques are increasingly applied across various fields, including epidemiological (Shah et al., 2014), hydro-meteorological (Jing et al., 2022), marketing and finance (Huang et al., 2022), education studies (Finch et al., 2016), etc. Unlike traditional statistical approaches, machine learning approaches do not require distributional assumptions about the data, enhancing their versatility. Numerous studies, including those by Stekhoven and Bühlmann (2012), Rizvi et al. (2023), and Tang and Ishwaran (2017), have demonstrated the effectiveness of these approaches in handling missing data. Specifically, Tang and Ishwaran (2017) observed that random forest imputation is effective in situations with moderate to high levels of missingness, including some cases of  MNAR data. Machine learning approaches seem to be effective in handling MNAR data. However, the use of these approaches in longitudinal research, particularly for addressing MAR and MNAR data, remains underexplored.

Typically, machine learning approaches for imputation use single imputation approaches, filling only one value for each missing observation. However, single imputation approaches do not consider the uncertainty of missing data, potentially leading to inefficient estimates (Enders, 2010). In contrast, multiple imputation approaches tackle this issue by creating several complete datasets. It estimates and imputes missing values multiple times and then consolidates these estimates into a single, more reliable result. This process acknowledges and incorporates the uncertainty associated with missing values, thereby enhancing the robustness of statistical analyses.

To adopt multiple imputation approaches, Van Buuren (2012) proposed Multivariate Imputation by Chained Equations (MICE), providing various imputation approaches such as predictive mean matching and linear regression. Doove, Van Buuren, and Dusseldorp (2014) expanded on this by incorporating machine learning techniques into MICE, combining the advantages of machine learning and multiple imputations. This integration led to more accurate and reliable missing data imputations, as Finch et al. (2016), Laqueur et al. (2022), and Shah et al. (2014) demonstrated. Nonetheless, there is still a relative scarcity of research evaluating the performance of machine learning approaches in both single and multiple imputation contexts (Stekhoven \& Bühlmann, 2012; Tang \& Tong, 2023) and comparing their performance with traditional techniques such as FIML and TSRE. 

In sum, despite the increasing use of machine learning approaches in handling missing data, there remains a notable gap in comprehensive research regarding their effectiveness and applicability, especially in longitudinal studies. Addressing this gap, this article is dedicated to assessing and comparing the performance of traditional missing data analytical techniques and machine learning imputation approaches, both single and multiple, in the context of longitudinal research. The comparison will consider various data distributions and missing data scenarios. This study aims to offer a detailed, quantifiable evaluation of these approaches in handling nonnormal ignorable and nonignorable missing data, ultimately providing practical guidelines for researchers conducting longitudinal studies.  Specifically, the next two sections will introduce traditional and machine learning approaches for missing data analysis, respectively, including two traditional approaches, two machine learning approaches by single imputation, and two machine learning approaches by multiple imputation.  Thereafter, a Monte Carlo simulation study will be conducted to evaluate the performance of the six approaches. Then, a real longitudinal data example with the National Longitudinal Survey of Youth 1997 Cohort data is provided to illustrate the application of the six approaches. Last, the article discusses the simulation results and provides practical recommendations for substantive researchers.

\section{Traditional Approaches for Missing Data Analysis}
In this section, we briefly introduce two prevalent traditional approaches for missing data analysis: Full Information Maximum Likelihood and the Two-Stage Robust Estimation. These approaches will be evaluated and compared in the simulation study.
\subsection{Full Information Maximum Likelihood (FIML)}
FIML, also known as normal-distribution-based maximum likelihood, is one of the most frequently used techniques for estimating model parameters in the presence of missing data. This technique has been implemented in popular SEM software like Mplus and lavaan (Asparouhov \& Muthen, 2007; Rosseel, 2012).  Conceptually grounded in maximum likelihood estimation principles, FIML estimates a likelihood function for each case in the data, measuring the discrepancy between observed data and the model's predictions based on current parameter estimates using only the available observations for each individual, and then, it maximizes the overall likelihood function that sums individual likelihoods (Raykov, 2005; Schafer \& Graham, 2002; Tang et al., 2024; Wen et al., 2018; Shin et al., 2017).  

Rather than directly imputing missing values, FIML estimates model parameters and standard errors using all available data, which functions like an imputation machine adjusting the model parameters based on the observed scores to infer the location of unseen data points (Enders,  2023). This approach typically yields more accurate parameter estimates than convenient approaches such as listwise deletion, pairwise deletion, and imputation approaches based on similar response patterns (Enders \& Bandalos, 2001). However, its reliance on the assumption of multivariate normality can be a limitation; violation of this normality assumption may lead to biased estimates.

In practice, longitudinal data are often nonnormally distributed (Cain et al., 2017). The presence of missing data makes it more difficult to detect the nonnormality (Yuan, Lambert, \& Fouladi, 2004). The observed data in an incomplete dataset may possess significant skewness and kurtosis even when the population distribution is normal. Likewise, the observed data could pass a skewness or kurtosis test when the population distribution is, in fact, nonnormal.  Tong et al. (2014) found that in longitudinal studies with nonnormal missing data, traditional chi-squared test statistics under FIML generally fail to reflect the nominal chi-square distribution accurately, so the test cannot maintain an appropriate Type I error rate.

\subsection{Two-Stage Robust Estimation (TSRE)}
TSRE is an advanced statistical technique for addressing missing data and potential nonnormality in the SEM framework. Similar to FIML, TSRE does not impute missing data directly. However, it can produce more robust and reliable parameter estimates when data violate the assumption of multivariate normality (Yuan et al., 2015).  

This technique has two stages. The first stage aims to obtain robust estimates for the mean vector and covariance matrix by downweighting potential outliers that are far away from the center of the majority of the data. This is accomplished through robust M-estimators (Yuan \& Zhang, 2012). The robustness of these estimators makes them particularly useful in data with nonnormal distributions or outliers. In the second stage, the SEM model is applied to the robust mean vector and covariance matrix obtained from the first stage. SEM model parameters, standard errors, and associated test statistics are estimated by minimizing a fit function and quantifying the discrepancy between the model-implied and the robustly estimated means and covariance matrices (Tong et al., 2014). By breaking down the estimation process into these two stages, TSRE effectively addresses the challenges posed by nonnormal and incomplete data, improving the accuracy and reliability of SEM analyses in the presence of missing data.

\section{Machine Learning Approaches for Missing Data Analysis}
In this section, we introduce frequently used machine learning approaches for missing data imputation, including single imputation techniques based on K-Nearest Neighbors, Classification and Regression Tree (CART), and Random Forest (RF), as well as multiple imputation approaches using CART and RF.
\subsection{K-Nearest Neighbors (KNN)} 
KNN, a supervised machine learning algorithm proposed by Fix and Hodges (1952), is used for classification and regression tasks. With the principle of KNN being that data points with similar characteristics are likely to yield similar values, the detailed procedure of KNN is summarized into the following three steps. The first step is to choose the number of 'k',  meaning that 'k' nearest neighbors will be used to infer the value for a missing observation.  The second step is to identify the 'k' observations that are closest to a missing observation, using similarity metrics such as Euclidean or Minkowski distance (Zhang, 2012). The third step is to estimate the missing value based on these observations. 

For categorical data, the missing value is assigned to the category most frequently observed among its k neighbors. For instance, Figure 1 shows that the observations are divided into classes A and B according to Features 1 and 2. If $k=3$, three nearest neighbors for the unobserved data point are identified, two in class A and one in class B. As a result, the estimated missing data is in class A. For continuous data, missing values are estimated as the mean or weighted mean of the k-nearest observations. This approach has been implemented in the \textit{VIM} (Visualization and Imputation of Missing Values) \textit{R} package developed by Kowarik and Templ (2016), using an extension of Gower's distance for prediction (Gower, 1971). 
%\centerline{[ Insert Figure 1 here ]}
\begin{figure}
	\centering
		\includegraphics[width=15cm,height=12cm]{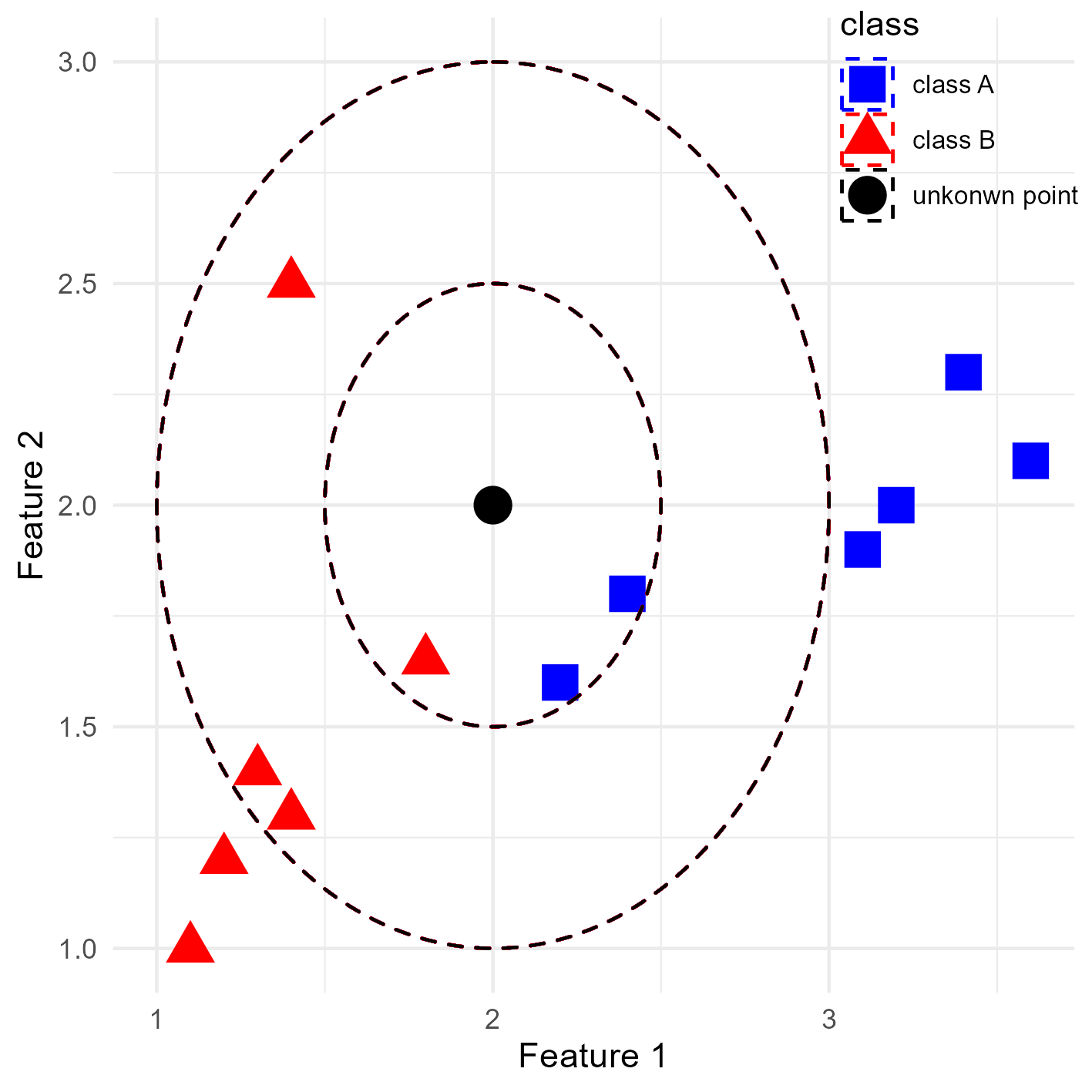}
	      \caption{An example of KNN imputation}  
       \label{fig:image1}
\end{figure}

The advantages of KNN imputation include its simplicity and intuitivity, which provide a clear visualization of its process. It also does not rely on the distributional assumption of the data and can handle non-linear relationships within high-dimensional data (Jönsson \& Wohlin,  2006). Moreover, it is robust to outliers because it relies on local information around missing values, often generating more accurate estimates than approaches using global information, like mean or median imputations (Batista et al., 2002).  

Despite being a single imputation approach, KNN has been found that it could handle missing data to reach the accuracy of the complete data with a low accuracy difference for categorical and continuous data (Jonsson \& Wohlin, 2004; Pujianto et al., 2019). Furthermore, it could generate more accurate estimates than imputation approaches based on the mean or mode (Parr et al., 2008; Ribeiro \& Freitas, 2019) and has been successfully applied to imputing missing data in longitudinal studies (Sania et al., 2021).

However, the performance of KNN imputation heavily depends on the choice of 'k'. A larger 'k' may incorporate less similar neighbors, potentially reducing the accuracy of the imputation. Conversely, a smaller 'k' may cause the imputation to be overly sensitive to local data variations. For instance, as illustrated in Figure 1, if 'k' is set to 3, the unknown data point has three nearest neighbors, with two belonging to class A and one to class B. As a result, the estimated unknown point is in class A.  If 'k' is increased to 6, the unknown point is surrounded by six nearest neighbors, now with a majority of four in class B and only two in class A, thereby reclassifying the unknown point to class B. Therefore, varying 'k' from 3 to 6 could shift the classification of the unknown point from class A to class B, demonstrating the sensitivity of KNN imputation to the choice of 'k' and the surrounding class distribution.

\subsection{Classification and Regression Tree (CART)}
The CART method, proposed by Breiman et al. (1984), is a decision tree-based supervised machine learning approach for classification and regression tasks. The principle of CART involves splitting the data into homogeneous subsets based on certain conditions of data features (Creel \& Krotki, 2006; Dagdoug et al., 2023, Serang \& Sears, 2021). These conditions manifest as decision nodes, where the data is divided into two branches based on each node's criterion (Stegmann et al., 2018). Note that the number of decision nodes should be predetermined. Then, these multiple data splits are combined to get an outcome called leaves.  For example, suppose we predict students' final exam scores (A) based on students' homework performance (B) and mid-term exam scores (C). Our data includes students' homework, mid-term exam, and final exam scores, ranging from 0 to 10. The CART algorithm evaluates and finds that splitting by "homework score (B) > 5" first minimizes the variance within the resulting subsets. Then, CART continues to split the data based on "mid-term exam scores (C) > 6" to make the subsets' variance minimal. Last, the outcomes are the average scores of each subset, as depicted in Figure 2. 
%\centerline{[ Insert Figure 2 here ]}
 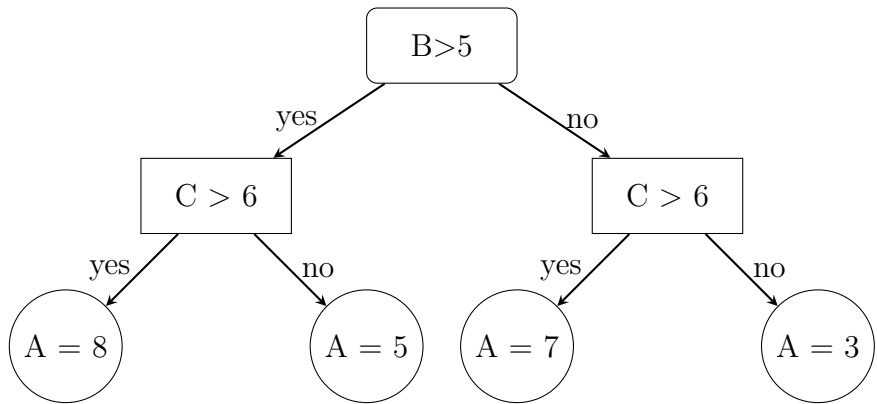
\begin{figure}
	\centering
	\vspace{0cm}
\pagestyle{empty} 
\tikzstyle{startstop} = [rectangle,rounded corners, minimum width=2cm,minimum height=1cm,text centered, draw=black]
\tikzstyle{process} = [circle,minimum size=1.5cm,text centered,draw=black]
\tikzstyle{decision} = [rectangle,minimum width=2cm,minimum height=1cm,text centered,draw=black]
\tikzstyle{arrow} = [thick,->,>=stealth]
\begin{tikzpicture}[node distance=1cm]
\node (start) [startstop,xshift=5cm] {B>5};

\node (decision1a) [decision,below of=start,yshift=-1cm,xshift=-3cm] {C > 6};
\node (decision1b) [decision,below of=start,yshift=-1cm,xshift=3cm] {C > 6};
\draw [arrow] (start) -- node[anchor=east] {yes} (decision1a);
\draw [arrow] (start) -- node[anchor=west] {no} (decision1b);

\node (process1a) [process,below of=decision1a,yshift=-1cm,xshift=-2cm] {A = 8};
\node (process1b) [process,below of=decision1a,yshift=-1cm,xshift=2cm] {A = 5};
\draw [arrow] (decision1a) -- node[anchor=east] {yes} (process1a);
\draw [arrow] (decision1a) -- node[anchor=west] {no} (process1b);

\node (process2a) [process,below of=decision1b,yshift=-1cm,xshift=-2cm] {A = 7};
\node (process2b) [process,below of=decision1b,yshift=-1cm,xshift=2cm] {A = 3};
\draw [arrow] (decision1b) -- node[anchor=east] {yes} (process2a);
\draw [arrow] (decision1b) -- node[anchor=west] {no} (process2b);
\end{tikzpicture}
\vspace{0cm}
	\caption{An example of a decision tree}
 \label{fig:image2}
\end{figure}

The CART can be used to impute missing data by creating a decision tree using observed data. Then, the decision tree is applied to predict missing values. For example, given the observed values of homework score B and mid-term exam score C, the missing value in the final exam A can be estimated based on the decision tree using the mean of the observations in the terminal node. This imputation provides a clear insight into the decision-making process, making it interpretable and transparent (Dagdoug et al., 2023).  Furthermore,  a key advantage of CART is that it can handle both categorical and continuous data, where the differences between the algorithms lie solely in the criteria used for data splitting and tree pruning methods (Loh, 2011). It also can  handle non-linear relationships within the original data without requiring data transformation.  However, this approach may overfit the data when the number of nodes is large, leading to poor generalization performance on predicted data (Schaffer, 1993).

While many studies have explored CART for missing data, they primarily used a single imputation approach (Creel \& Krotki, 2006; Conversano \& Siciliano, 2009; Dagdoug et al., 2023).  However, the single imputation approach cannot adequately account for the uncertainty associated with the missing values. It thus can distort the distribution of the data and make the results of subsequent analyses less efficient. Therefore, Doove et al. (2014) proposed a multiple imputation approach by CART, called 'micecart', and implemented the approach in the \textit{R} packages \textit{mice} (Van Buuren, 2018). 

In a longitudinal setting with observations at $T$ time points, $\textbf{Y} = ( \textbf{y}_{1}, ..., \textbf{y}_{T} )$,  the procedures of micecart are described as follows: 

(1)  For each variable $\textbf{y}_t$ with missing data, create initial imputations, which are randomly drawn from the observed values of $\textbf{y}_t$ (denoted as  $\textbf{y}^{obs}_t$). The current data matrix after imputation is denoted as $ \textbf{Z}$. Note that $ \textbf{Z}$ may also contain covariates that are completely observed.

(2) For each variable with missing values $\textbf{y}_t$, fit a CART using $\textbf{y}_t$ as the outcome and  $\textbf{Z}$ as predictors. For each subject in $\textbf{y}_t$, determine the terminal node based on the CART model. Then, impute the missing values by randomly selecting from the observed values in that terminal node.

(3) Repeat step 2 for multiple iterations. Each iteration should use the imputed values from the previous iteration as the updated $ \textbf{Z}$ matrix to fit a new CART model for each variable with missing data.  The number of iterations can be predetermined based on the convergence criteria.

(4) Perform steps 1-3 multiple times, with each time creating a separate imputed dataset. The number of imputations (\textit{M}) will result in \textit{M} complete datasets.

These datasets can be analyzed separately with results pooled according to Rubin's rules (Javadi et al., 2021), such as averaging parameter estimates across the imputed datasets. This imputation approach provides a robust solution to the missing data problem while accounting for the uncertainty inherent in the imputation process.

Numerous studies have demonstrated the effectiveness of the micecart approach in addressing missing data. Shah et al. (2014) discovered that micecart could yield unbiased parameter estimates and appropriate confidence intervals. Wongkamthong and Akande (2023) found that micecart generally outperformed the other multiple imputation approaches, such as MICE using multinomial logistic regression models. Wang et al. (2022) reported that micecart significantly outperformed deep learning imputation approaches in terms of bias, mean squared error, and coverage in various realistic settings.

\subsection{Random Forest (RF)}
RF, proposed by Breiman (2001) and Ho (1995), is another machine learning approach for classification and regression. In the Random Forest algorithm, multiple subsets of data are first created using bootstraps, sampling with replacement. Then, a decision tree is built from each subset. The final output is chosen by combining all the decisions. For instance, as illustrated in Figure 3, the process of RF involves creating two trees. In classification, the outcome is the class most frequently selected by the trees, while for regression, it is the mean or average prediction from all trees.
%\centerline{[ Insert Figure 3 here ]}
\begin{figure}
	\centering
	\vspace{0cm}
\pagestyle{empty} 
\tikzstyle{startstop} = [rectangle,rounded corners, minimum width=2cm,minimum height=1cm,text centered, draw=black]
\tikzstyle{tree} = [rectangle,rounded corners, minimum width=1cm,minimum height=0.8cm,text centered, draw=black]
\tikzstyle{leaf} = [circle,minimum size=1cm,text centered,draw=black]
\tikzstyle{decision} = [rectangle,minimum width=2cm,minimum height=1cm,text centered,draw=black]
\tikzstyle{subnode} = [rectangle,minimum width=1cm,minimum height=0.8cm,text centered,draw=black]
\tikzstyle{arrow} = [thick,->,>=stealth]
\tikzstyle{arrow2} = [thick,->,>=stealth,minimum width=1cm,minimum height=0.8cm]
\begin{tikzpicture}[node distance=1cm]
\node (start) [startstop,xshift=5cm] {All Data};
%subset
\node (decision1a) [decision,below of=start,yshift=-0.5cm,xshift=-3cm] {subset};
\node (decision1b) [decision,below of=start,yshift=-0.5cm,xshift=3cm] {subset};
\draw [arrow] (start) -- node[anchor=east] {} (decision1a);
\draw [arrow] (start) -- node[anchor=west] {} (decision1b);

%tree1
\node (subtree1) [tree,below of=decision1a,yshift=-0.5cm] { };
\node (node1a) [subnode,below of=subtree1,yshift=-0.5cm,xshift=-1.55cm] {};
\node (node1b) [subnode,below of=subtree1,yshift=-0.5cm,xshift=1.55cm] {};
\draw [arrow] (subtree1) -- node[anchor=east] {} (node1a);
\draw [arrow] (subtree1) -- node[anchor=west] {} (node1b);
%leave
\node (leaf1a) [leaf,below of=node1a,yshift=-0.5cm,xshift=-0.9cm] {};
\node (leaf1b) [leaf,below of=node1a,yshift=-0.5cm,xshift=0.9cm] {};
\draw [arrow] (node1a) -- node[anchor=east] {} (leaf1a);
\draw [arrow] (node1a) -- node[anchor=west] {} (leaf1b);

\node (leaf1c) [leaf,below of=node1b,yshift=-0.5cm,xshift=-0.9cm] {};
\node (leaf1d) [leaf,below of=node1b,yshift=-0.5cm,xshift=0.9cm] {};
\draw [arrow] (node1b) -- node[anchor=east] {} (leaf1c);
\draw [arrow] (node1b) -- node[anchor=west] {} (leaf1d);

%tree2
\node (subtree2) [tree,below of=decision1b,yshift=-0.5cm] { };
\node (node1c) [subnode,below of=subtree2,yshift=-0.5cm,xshift=-1.55cm] {};
\node (node1d) [subnode,below of=subtree2,yshift=-0.5cm,xshift=1.55cm] {};
\draw [arrow] (subtree2) -- node[anchor=east] {} (node1c);
\draw [arrow] (subtree2) -- node[anchor=west] {} (node1d);

%leave
\node (leaf2a) [leaf,below of=node1c,yshift=-0.5cm,xshift=-0.9cm] {};
\node (leaf2b) [leaf,below of=node1c,yshift=-0.5cm,xshift=0.9cm] {};
\draw [arrow] (node1c) -- node[anchor=east] {} (leaf2a);
\draw [arrow] (node1c) -- node[anchor=west] {} (leaf2b);

\node (leaf2c) [leaf,below of=node1d,yshift=-0.5cm,xshift=-0.9cm] {};
\node (leaf2d) [leaf,below of=node1d,yshift=-0.5cm,xshift=0.9cm] {};
\draw [arrow] (node1d) -- node[anchor=east] {} (leaf2c);
\draw [arrow] (node1d) -- node[anchor=west] {} (leaf2d);
%decision
\node (classA) [decision,below of=leaf1c,yshift=-0.5cm,xshift=-0.4cm] {Decision X};
\node (classB) [decision,below of=leaf2c,yshift=-0.5cm,xshift=-0.4cm] {Decision X};
\node (vote) [decision,below of=classA,yshift=-0.5cm,xshift=3.1cm] {majority descision};
\draw [arrow] (classA) -- node[anchor=east] {} (vote);
\draw [arrow] (classB) -- node[anchor=west] {} (vote);
%final
\node (final) [startstop,below of=vote,yshift=-0.8cm] {finial descision};
\draw [arrow] (vote) -- (final);
\end{tikzpicture}
\vspace{0cm}
	\caption{A diagram of a random forest}
	\label{fig:image3}
\end{figure}
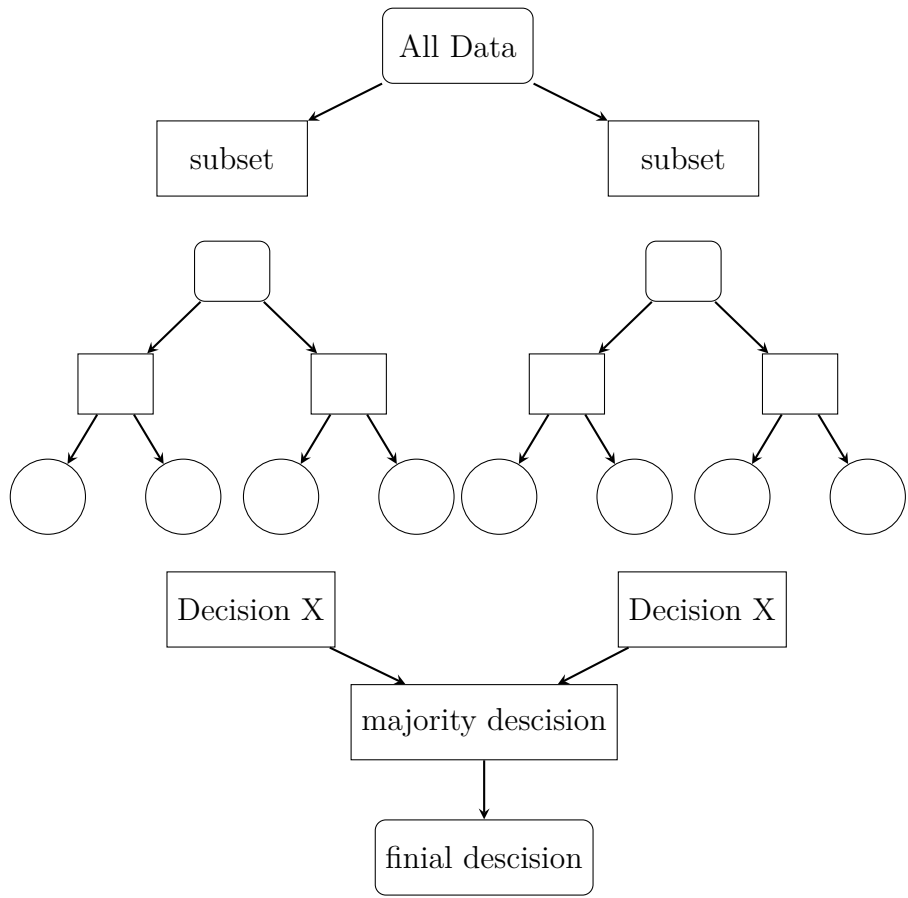

Unlike CART, random forests create multiple trees instead of only one. By creating many trees, RF reduces the probability of overfitting. Moreover, RF does not require distributional assumptions about the data and can efficiently handle various data types, including numerical, categorical, or mixed, without preliminary transformations. It is particularly good at managing high-dimensional data and capturing complex interactions and nonlinear relationships (Stekhoven \& Bühlmann, 2012).

Random Forest has been widely applied in missing data imputation (Ishwaran et al., 2008; Stekhoven \& Bühlmann, 2012). One of the most popular random forest imputation approaches is missForest, implemented in the \textit{missForest} \textit{R} package (Stekhoven \& Bühlmann, 2012). The procedures of missForest are described as follows: 

 (1) Fill in missing values with initial guesses. 
 
 (2) Fit a Random Forest model to the data. For each variable with missing data, treat that variable as the response and all other variables as predictors.  Use this Random Forest model to predict the missing values.

 (3) After the first round of imputation, missForest repeats the imputation process. It re-fits the Random Forest models with the newly imputed data and updates the imputations.

(4) After each iteration, missForest checks for convergence by comparing the newly imputed data matrix with the one from the previous iteration. If their difference is smaller than a predefined threshold, the algorithm terminates and reports convergence. 

(5)  Once the algorithm converges or reaches a maximum number of iterations, it outputs the imputed dataset.

Wei et al. (2018) reported that missForest outperformed seven other imputation approaches, such as mean, median, and KNN for MCAR and MAR  data. For time series data, missForest can effectively restore the missing information (Zhang et al., 2021).  

Some researchers considered missForest a multiple imputation approach because it averages over many unpruned classification or regression trees to get final imputed values (Stekhoven \& Bühlmann, 2012; Zhang et al., 2021).  Although missForest performs the imputation process multiple times to improve accuracy, each missing value is actually imputed only once in the final dataset (Hong \& Lynn, 2020).  Therefore, missForest is still a single imputation approach.   

To overcome the limitation of the single imputation approach, Doove et al. (2014) developed a multiple imputation approach using random forests based on multivariate imputation by chained equations, called miceForest and implemented in the mice R package. The process of miceForest is similar to micecart, where CART is changed into FR in the second and third steps. Research indicates this approach performed well, especially when interaction effects exist in data (Doove et al., 2014). Finch et al. (2016) found that miceForest could generate more accurate estimates than the RF alone. While some studies have shown the efficacy of missForest (Stekhoven \& Bühlmann, 2012; Smith et al., 2021; Tang \& Ishwaran, 2017), comparisons of these approaches in longitudinal studies are not available. Therefore, in this study, we will explore these two imputation approaches in the context of longitudinal data.

\section{A Simulation to Compare the Performance of Traditional and Machine Learning Approaches}
\subsection{Study design}
A Monte Carlo simulation study is conducted to evaluate the performance of traditional and machine learning missing data analytical techniques in longitudinal research.  Since growth curve models (GCMs) are commonly used in analyzing longitudinal data (Drake et al., 2020; Itzchakov et al., 2023) with the ability to directly investigate intraindividual change over time and interindividual differences in intraindividual change, we study the performance of traditional and machine learning approaches in the context of growth curve modeling. 

Consider a study with a cohort of $N$ individuals, each measured $T$ times. For the $i$th individual, let $\textbf{y}_i = (y_{i1}, ..., y_{iT} )$ denote their observed scores, with $y_{it}$ representing the score at the $t$th measurement occasion ($t= 1, ..., T$). A typical unconditional GCM can be expressed as follows:

\begin{equation}
 \begin{aligned}
 & \textbf{y}_i = \mathbf{\Lambda}\textbf{b}_i + \textbf{e}_{i},	\\
 & \textbf{b}_i = \mathbf{\beta} + \textbf{u}_i,
 \end{aligned}
\end{equation}
where $\mathbf{\Lambda}$ is a $T \times q$ factor loading matrix that captures the growth trajectory shapes. The vector $\textbf{b}_i$ (a $q \times 1$ vector) represents the individual-specific random effects, while $\textbf{e}_{i}$ denotes intraindividual measurement errors. The random effects $\textbf{b}_i$ vary among individuals, and their mean, $\mathbf{\beta}$, corresponds to fixed effects for the entire population. The residual vector $\mathbf{u}_i$ denotes the random components of $\textbf{b}_i$. 

It is typically assumed that both $\textbf{e}_{i}$ and $\textbf{u}_i$ are multivariate normally (MN) distributed, such that:
\begin{equation}
 \begin{aligned}
 & \textbf{e}_{i} \sim \mathcal{MN}_T(\textbf{0},\mathbf{\Phi}),\\
 & \textbf{u}_i \sim \mathcal{MN}_q(\textbf{0},\mathbf{\Psi}).
 \end{aligned}
\end{equation}

In these equations, the subscript in the MN distribution indicates the dimensionality of the random vector. $\mathbf{\Phi}$, a $T \times T$ covariance matrix for $e_i$, is often assumed to be a diagonal matrix ($\Phi = \sigma^2_eI$), suggesting that intraindividual measurement errors are consistent in variance and independent across different measurement occasions.

Following  the empirical data analysis results in Tong, Zhang, and Yuan (2014), our data are generated from a linear growth curve model with four measurement occasions ( $T$ = 4).  As Figure 4 shows, in the population model, the fixed effects of the latent intercept and slope are 6 and 2 ($\beta=(\beta_L, \beta_S)'= (6, 2)'$), respectively. The variance of the latent intercept is 1 ($\sigma_L^2$ = 1), the variance of the latent slope is 1 ($\sigma_S^2$ = 1), and the correlation between the latent intercept and slope is 0 ($\sigma_{LS}$ = 0). The variance of the intraindividual measurement errors is also set at 1 ($\sigma_e^2$ = 1). 

%\centerline{[ Insert Figure 4 here ]}

\begin{figure}
	\centering
		\includegraphics[width=7.5cm,height=8cm]{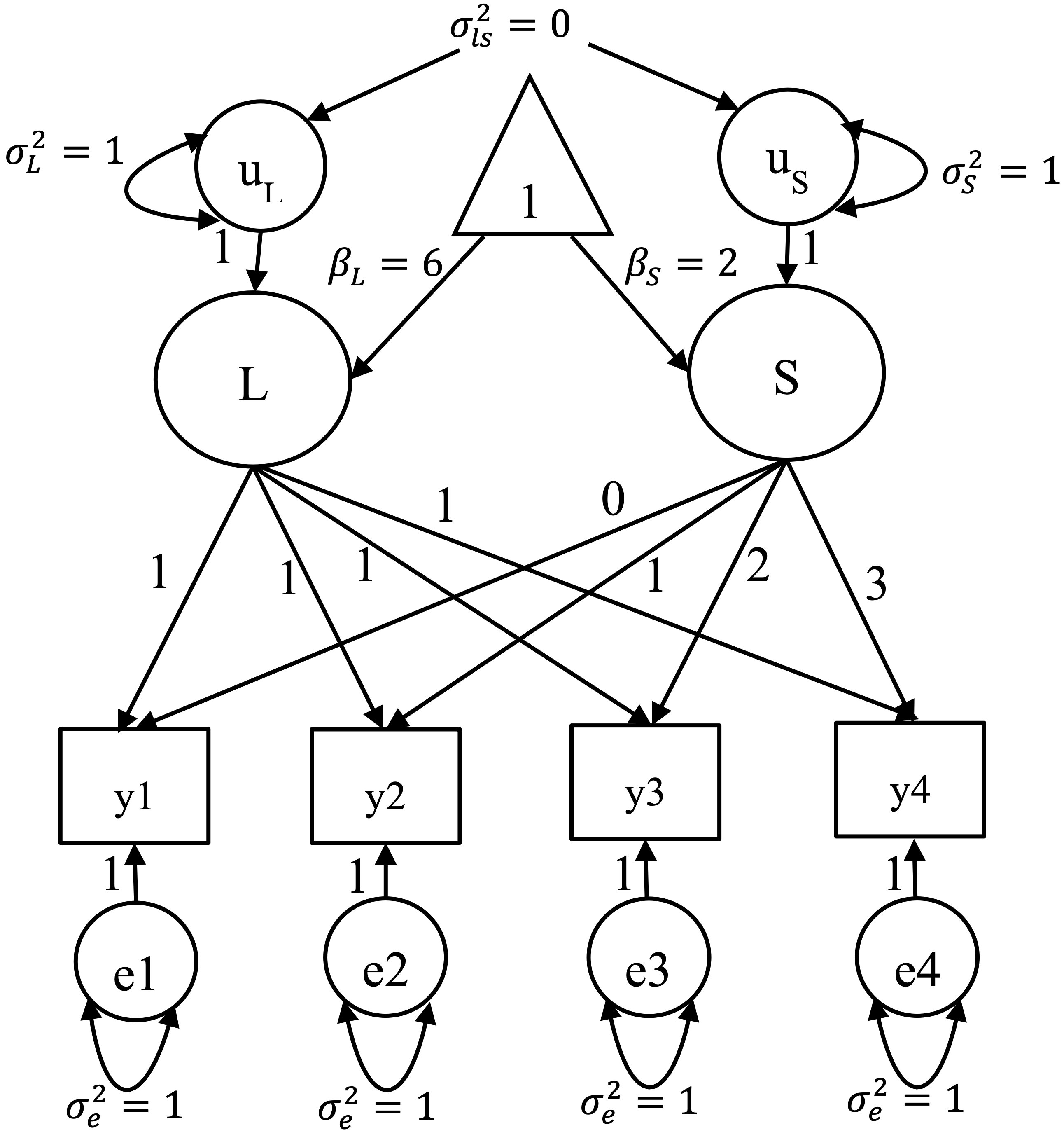}
	      \caption{Path diagram of a growth curve model.}  
       \label{fig:image4}
\end{figure}

\subsection{Manipulated factors}

In this simulation, four potentially influential factors are manipulated as follows.

(1) Sample size: 100, 200, 500, 1000, and 5000. Small (100), medium (200), and large (500) sample sizes are typically used in growth curve modeling (e.g., Tong, Zhang, \& Zhou, 2021).  Two larger sample sizes, 1000 and 5000, are also incorporated into our simulation study as they are commonly seen in studies utilizing machine learning imputation methods (e.g., Jerez et al., 2010). 

(2) Missing data mechanism: MAR and MNAR.  For the MAR mechanism, observations $y_{i2}$, $y_{i3},$ and $y_{i4}$ are manipulated to be missing when $y_{i1} > c_1$, where $c_1$ is a cutoff value determined by the missingness rate. If $y_{i2}$ is observed, then observations $y_{i3}$ and $y_{i4}$ are missing when $y_{i2} > c_2$, and so forth (Tong, Zhang, \& Yuan, 2014). Thus, all the missing values are MAR, as the missingness only depends on observed scores of $\textbf{y}_i$ in the model. For the MNAR mechanism, an auxiliary variable ($Aux_i$) is generated for each subject, which is expressed by $Aux_i = r\times b_{is} + \epsilon_i$, where $\epsilon_i$  following a standard normal distribution  $\epsilon_i \sim \mathcal{N}{(0,1)}$, $b_{is}$ is the random slope for the $i$th individual, and $r$ is the regression coefficient of the latent slope and is set in our simulation study to maintain a high correlation (=0.8) between the auxiliary variable and the latent slope. If $Aux_i$ is larger than a given percentile $p_t$ (t =2, 3, and 4), $y$ is missing, where $p_t$ is determined by the missing data rate. The auxiliary variable is not included in the growth curve analysis and is related to the unobserved latent slope; thus, the missingness is related to an unobserved variable, leading to MNAR data. Note that MCAR is not evaluated in the simulation study because most missing data analytical techniques can effectively handle MCAR data. We mainly study MAR as an illustration of ignorable missing data and MNAR as an illustration of nonignorable missing data.

(3)  Rate of missingness ($mr$):  0\%, 5\%, 15\%, and 30\%.  Complete data with 0\% missingness serves as the baseline for evaluating the effect of the missing data rate. We also manipulated 5\%, 15\%, and 30\% missingness as small, medium, and large levels, respectively (Tong, Zhang, \& Zhou, 2021; Lei \& Shiverdecker, 2020). To obtain the desired missingness rate $mr$, for the MAR condition, $c_t$, where $(t = 1,..., T-1)$, are set as at these upper levels of $y_t$: $1-[2t/(T-1)]\times mr$. For MNAR condition, $p_t$, where $(t = 2, ..., T)$, are set as lower $(1-[2(t-1)/(T-1)]\times mr)th$ percentiles of this normal distribution $Aux_i\sim \mathcal{N}{(r\times b_{is},r^2+1)}$.

(4) Data distributions: normal distribution, lognormal distribution, Student's t distribution, and normal distribution with $5\%$ outliers. Different population distributions are generated by manipulating the distributions of intraindividual measurement errors $\textbf{e}_{i}$. The distributional levels are set as follows: normal distribution $\textbf{e}_{i}\sim N(0,\sigma^2_e)$, lognormal distribution $\textbf{e}_{i}\sim LN_{(0,1)}(0,\sigma^2_e)$,  t distribution $\textbf{e}_{i}\sim t_{(5)}{(0,\sigma^2_e)}$, normal distribution with $5\%$ outliers, where outliers are generated from a normal distribution with the same variance but a shifted mean $\textbf{e}_{i}\sim N(5,\sigma^2_e)$.  These levels provide various nonnormal data commonly seen in psychological studies, including the presence of outliers, large skewness, and kurtosis (Cain et al., 2017; Micceri, 1989).

Overall, there are 140 data conditions in this simulation, calculated as  $ 5 \times 2 \times (4-1) \times 4 + 5 \times 1 \times 4$. For each data condition, 500 datasets were generated and analyzed using the six missing data analytical approaches. The six approaches are traditional approaches, including FIML and TSRE, as well as the machine learning approaches by single imputation (KNN and missForest) and the machine learning approaches by multiple imputation (micecart, and miceForest). The R packages \textit{lavaan} and \textit{rsem} were used for missing data analysis with FIML (Rosseel, 2012) and TSRE (Yuan \& Zhang, 2011), respectively. The KNN imputation and single imputation by random forest (missForest) approaches were conducted using the \textit{VIM} and \textit{missForest} packages in \textit{R}, respectively (Stekhoven \& Bühlmann, 2012; Templ et al., 2022).  The multiple imputations by CART (micecart) and random forest (miceForest) were conducted using the \textit{mice} package in \textit{R} (Van Buuren \& Groothuis-Oudshoorn, 2011).  The number of multiple imputations in the simulation was set at 20 because previous studies have investigated the number of imputations required for multiple imputation and FIML to produce comparable results and recommended a minimum of 20 imputations per dataset (Graham et al., 2007).

\subsection{Hyperparameters}
Before conducting the simulation, it was necessary to set hyperparameters that guide the learning process in machine learning approaches. Specifically, for  KNN, CART, and RF, these hyperparameters include the number of trees, nodes, and nearest neighbors.  To optimize the performance of these approaches, we analyzed the datasets for each condition using various levels of these hyperparameters, as detailed in Table 1. 
%\centerline{[ Insert Table 1 here ]}
\begin{table}[ht]
	\centering
	\caption{The Levels of the Hyperparameters for the Machine Learning Approaches}
	\label{t1} 
	\begin{tabular}{rrrr}
		\hline
		approach& hyperparameter & Level & optimal \\ 
		\hline
		KNN& number of nearest neighbours & 5-18& 5 \\ 
		missForest & number of trees & 10, 50, 100 & 10\\ 
		miceForest  &  number of trees & 10, 50, 100 & 10 \\ 
            micecart  &  number of nodes & 5, 10, 15 & 5 \\ 
		\hline
	\end{tabular}
\end{table}

The optimal hyperparameter settings were determined based on their performance, which was evaluated using relative bias (RB) and mean squared error (MSE). These two evaluation criteria are further elaborated in the next subsection.  The results of these machine learning approaches with various hyperparameters are available on our \href{https://github.com/DandanTang0/Evaluation-of-Missing-Data-Analytical-Techniques-in-Longitudinal-Research}{GitHub} site. As Table 1 illustrates,  the following optimized hyperparameters were chosen: 5 nearest neighbors  for KNN, 10 trees for missForest, 10 trees for miceForest, and 5 nodes for micecart. Then, we analyzed the simulated data using traditional approaches and machine learning techniques with these optimized hyperparameters.

\subsection{Evaluation criterion}
The performance of traditional and machine learning analytical approaches was evaluated using the relative bias and mean squared error of the model parameter estimates.  The relative bias quantifies that the average estimate among 500 replicates differs from the true value of the parameter relative to the true value itself.  It can be expressed by $RB= 100\%\times\frac{\bar{\theta}-\theta}{\theta}$, where $\theta$ denotes a population parameter, and $\bar{\theta} = \frac{1}{500}\sum_{j=1}^{500}\hat{\theta}_j$, with $\hat{\theta}_j$  denoting a parameter estimate from the  $jth$ simulation replication,  $j = 1,...., 500$. When the population parameter is zero ($\theta = 0$),  RB simplifies to $RB= \bar{\theta}-\theta$.  Generally,  $|RB| < 10\%$ is considered an acceptable bias (Hoogland  \& Boomsma, 1998; Wen et al., 2019). 

The mean squared error measures the average squared difference between the estimated values and the actual value, which can be expressed by $MSE = \frac{1}{500}\sum_{j=1}^{500}(\hat{\theta}-\theta)^2$. MSE takes into account both the bias and the variance of the estimator, indicating how far the average estimate is from the true value as well as how spread out the estimates are. A lower MSE indicates that an estimator is both more accurate and precise.

\subsection{Results}
\subsubsection{Methods comparison}
The latent slope and the variance of the latent slope are often of primary interest to substantive researchers. Due to space limitations, only results for the estimated average of the variance of  latent slopes $\sigma_S^2$ are displayed in Figures 5 to 8. Results for other parameters show similar patterns in terms of the comparison among the six techniques and thus are put in the supplementary documents on our \href{https://github.com/DandanTang0/Evaluation-of-Missing-Data-Analytical-Techniques-in-Longitudinal-Research}{GitHub} site. According to simulation results for both MAR and MNAR data, in general, traditional approaches performed better than machine learning approaches, and approaches with multiple imputations performed better than the single imputation approaches in terms of the parameter estimation bias and mean squared error. Specifically, for MAR, when the data followed a normal distribution, a Student's t distribution, or a normal distribution with $5\%$ outliers, FIML and TSRE yielded the smallest parameter estimation bias, followed by micecart, which outperformed miceForest. The miceForest, in turn, was superior to missForest, which performed better than KNN. When the data were lognormally distributed, TSRE performed better than FIML in general, and they both provided less biased and more efficient parameter estimates than micecart, while other approaches performed similarly and were more biased and less efficient. For MNAR, the relative performance of these six approaches varied across different levels of sample size,  rate of missingness, and data distribution. 

Detailed results are presented in the following regarding the manipulated factors in the simulation study. 

\subsubsection{Sample size}  
Under the MAR mechanism, for normally distributed and Student's t distributed data, the performance of FIML and TSRE remained steady across different sample size conditions. As shown in Figures 5 and 7, both FIML and TSRE yielded acceptable bias in estimating the average of the variance of latent slopes. As the sample size increased, the other four approaches produced less biased and more efficient parameter estimates. However, these machine learning approaches only produced acceptable estimation bias when the missingness rate was small. With relatively large missing rates, increasing sample size did not help improve the performance of these machine learning approaches.  For data following lognormal distribution and normal distribution with outliers, the performance of FIML slightly deteriorated in estimating $\sigma_S^2$ as the sample size increased, while the other five techniques improved with larger sample sizes, which can be seen by comparing the subfigures in the second row of the figure in the top panel of Figures 6 and 8. When the missingness rate was small, all approaches produced acceptable bias.

Under the MNAR mechanism, the accuracy of the six techniques in estimating $\sigma_S^2$ improved with the increase in sample size,  as shown by comparing the subfigures in the first row of the figure in the top panel of Figures 5-8. Regardless of sample size, all techniques produced acceptable bias only when the missing data rate was small.  

\subsubsection{Rate of missingness} 
Under the MAR mechanism, the impact of the missingness rate on the performance of the six missing data analytical techniques in estimating average of the variance of latent slopes also varied across different data distributions. For normal distribution and Student's t distribution, FIML and TSRE had a stable performance for different levels of missingness rate, and both yielded acceptable bias (i.e., $|RB| < 10\%$). For example, for Student's t distribution, even when the missing data rate was 30\%, the relative bias for estimated $\sigma_S^2$ based on FIML and TSRE was smaller than 5\% for all sample size conditions, as shown in Figure 7. The machine learning approaches showed substantially decreased accuracy as the missing rate increased and only produced acceptable bias when the missing rate was small.  For lognormal distribution and normal distribution with outliers,  all the six missing data analytical techniques displayed decreased performance in estimating model parameters as the missing data rate increased and  produced acceptable bias only when the missing rate was small. For example, as shown in Figure 6, when the sample size was 1000, and the missing data rate was 2.5\%, the relative bias of estimated $\sigma_S^2$ was smaller than 10\%, but when the missing data rate increased to 30\%, the relative bias increased to larger than 20\%.

Under the MNAR mechanism, when the missing data rate was small, all techniques produced acceptable bias ($|RB| < 10\%$).  However, as the missingness rates increased, the estimation bias increased substantially. For example, even when data are normally distributed (see Figure 5), and the sample size was 1000, the relative bias for the average of the variance of latent slope estimates increased from lower than 10\% to larger than 30\% as the missingness rate increased from 2.5\% to 30\%. 

\subsubsection{Data distributions} 
Under the MAR mechanism, FIML and TSRE exhibited comparable or superior performance to the other four machine learning techniques in estimating the average of the variance of latent slopes for data following normal distribution, Student's t distribution, and normal distribution with $5\%$ outliers, which can be seen by examining the first subfigure in the second row of the figure in the top panel of Figures 5,  6, and 8. For lognormal distribution, TSRE showed the best performance, especially at higher missingness rates. For example, as shown in Figure 6, when the sample size was as large as 1000, and the missing data rate was 30\%, the MSE of the estimated $\sigma_S^2$ based on TSRE was smaller than 0.15, was larger than 0.15 based on FIML, and above 0.6 based on the machine learning approaches. 

Under the MNAR mechanism, FIML outperformed the other approaches in estimating model parameters for data following normal distribution, Student's t distribution, and normal distribution with $5\%$ outliers, which can be seen by examining the first row of the subfigures in Figures 5,  6, and 8. For lognormal distribution, FIML outperformed other approaches at relatively small sample sizes ( $N \leq 500$),  while missForest excelled at very large sample sizes and low missingness rates, as shown by the fifth subfigures in the first row of the subfigures in Figure 6. 

\subsection{Summary}
This simulation study assessed the performance of six missing data analytical techniques under various data conditions. For MAR missing data, TSRE was preferred. If the data are normally distributed, Student's t distributed, or normally distributed with $5\%$ outliers, FIML was comparable to TSRE. Machine learning methods and TSRE did not outperform FIML for handling MNAR missing data. However, missForest was preferable with lognormal distributed data at very large sample sizes (e.g., $n \geq 1000$) and low missingness rates.  

%\centerline{[ Insert Figure 5 here ]}
%\centerline{[ Insert Figure 6 here ]}
%\centerline{[ Insert Figure 7 here ]}
%\centerline{[ Insert Figure 8 here ]}

\begin{figure}[h]
    \centering
    \includegraphics[scale=0.1]{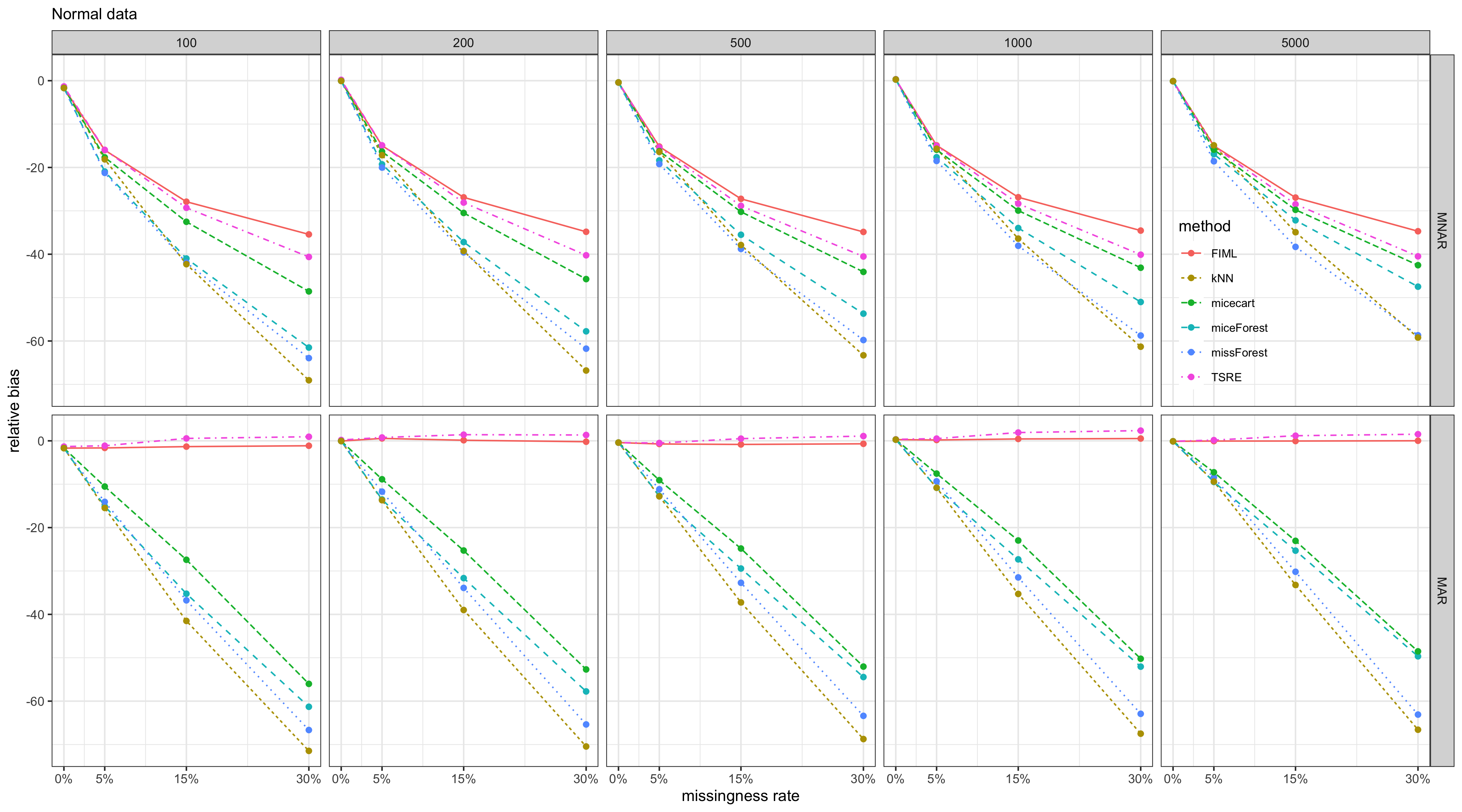} 
    
    \vspace{0.1cm} 
    
    \includegraphics[scale=0.1]{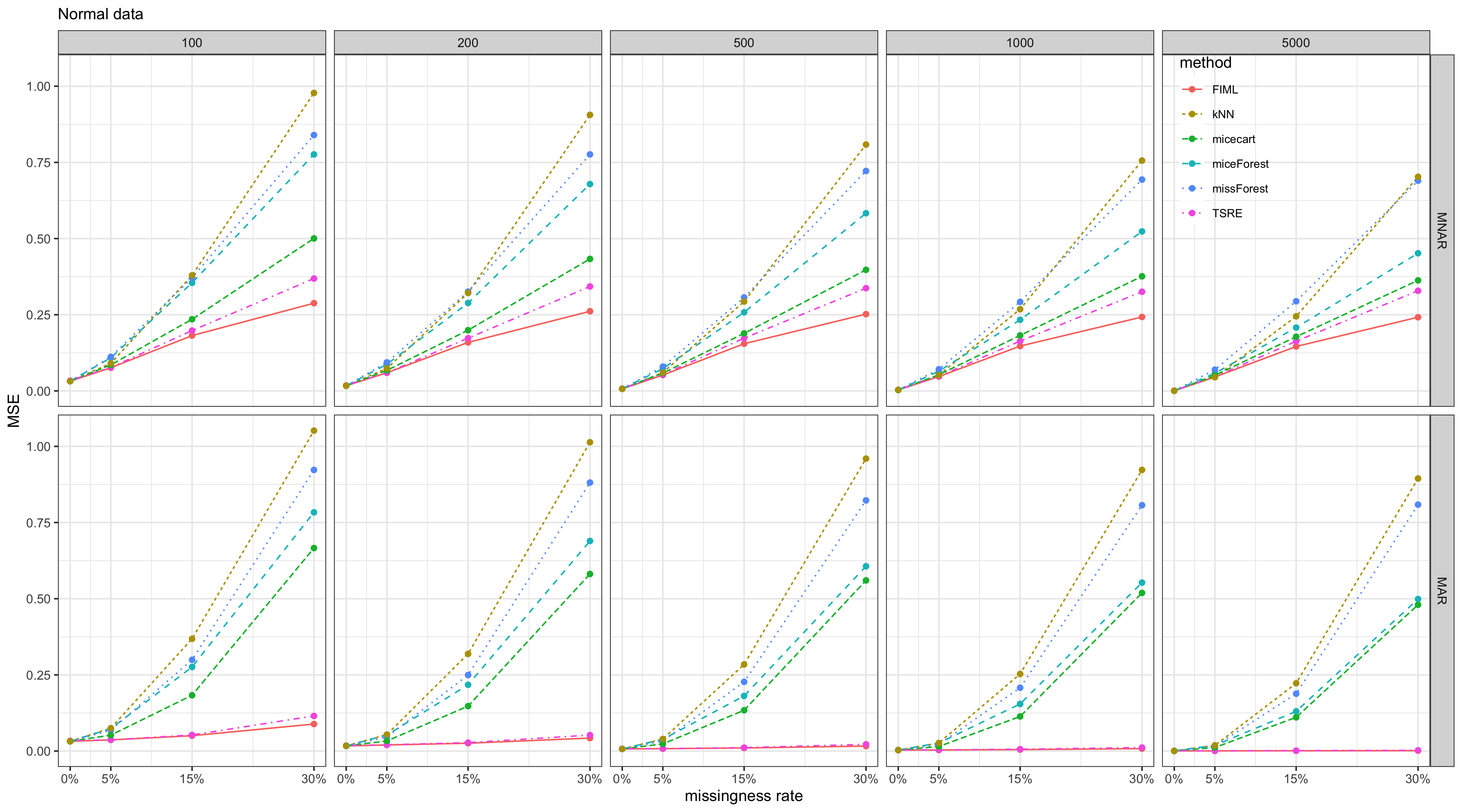} 
    \caption{The relative bias and mean squared error for the average of the variance of the latent slope estimates under the normal distribution} 
    \label{fig:image5}
\end{figure}

\begin{figure}[h]
    \centering
    \includegraphics[scale=0.1]{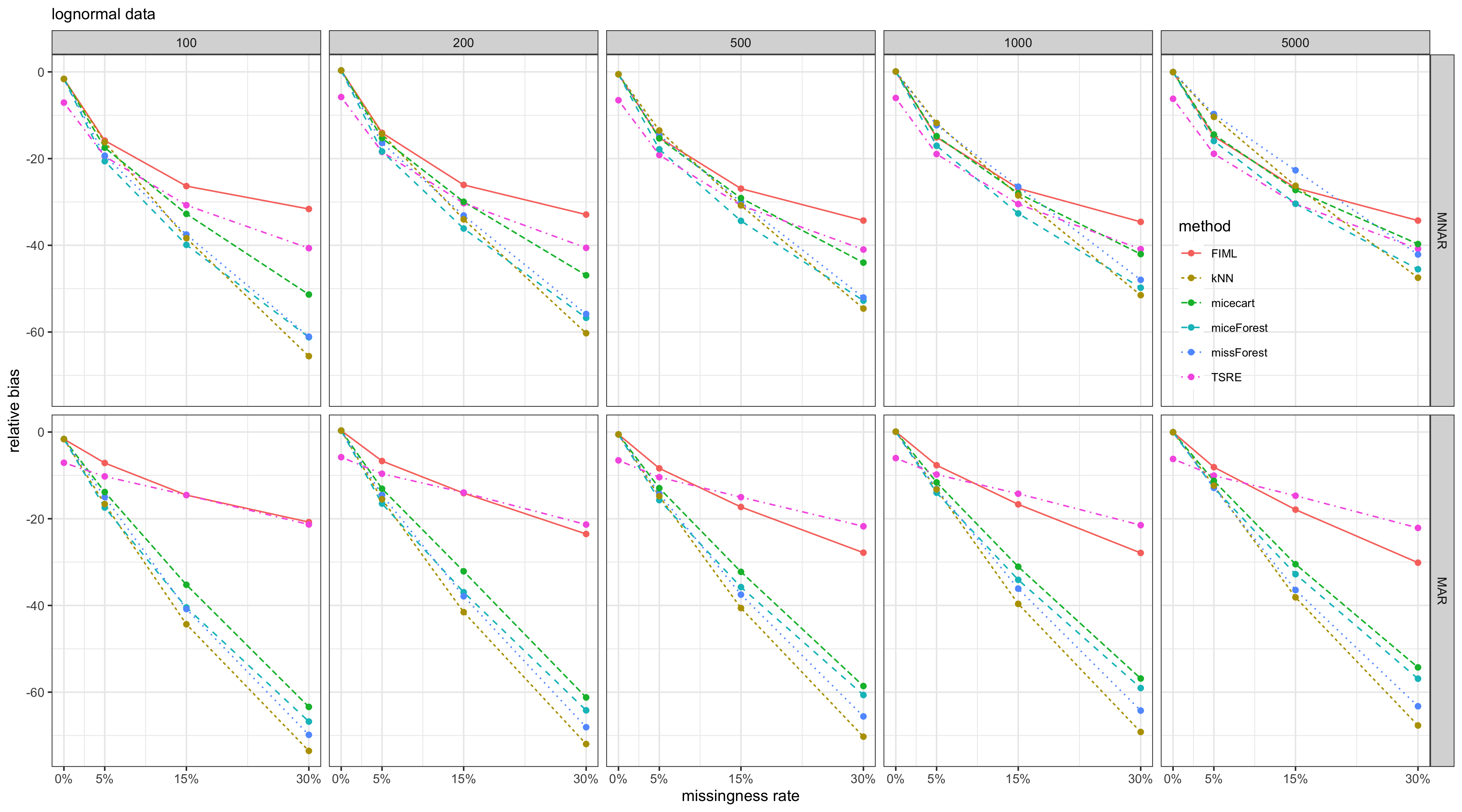} 
    
    \vspace{0.2cm} 
    
    \includegraphics[scale=0.1]{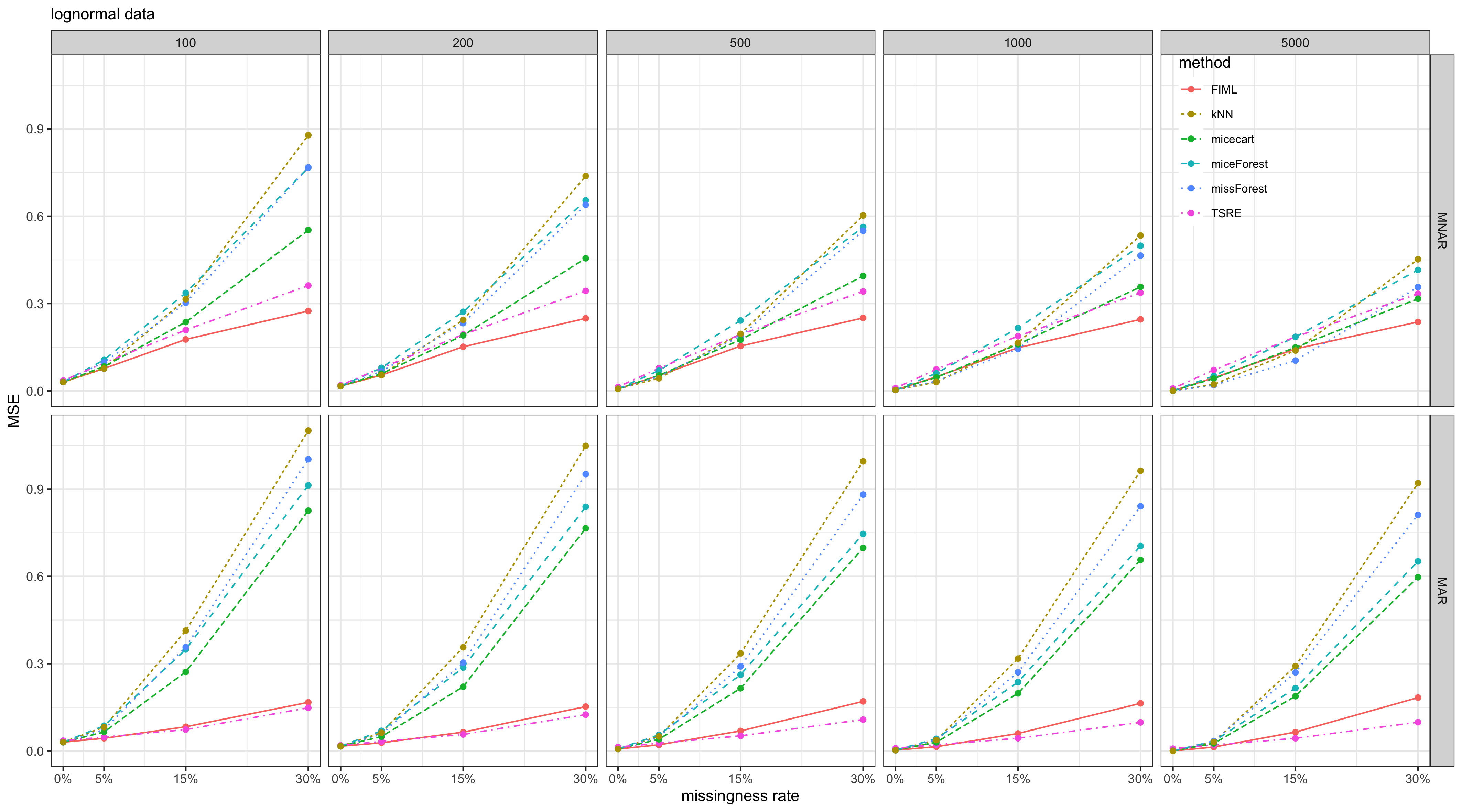} 
    \caption{The relative bias and mean squared error for the average of the variance of the latent slope estimates under the lognormal distribution} 
    \label{fig:image6}
\end{figure}

\begin{figure}[h]
    \centering
    \includegraphics[scale=0.1]{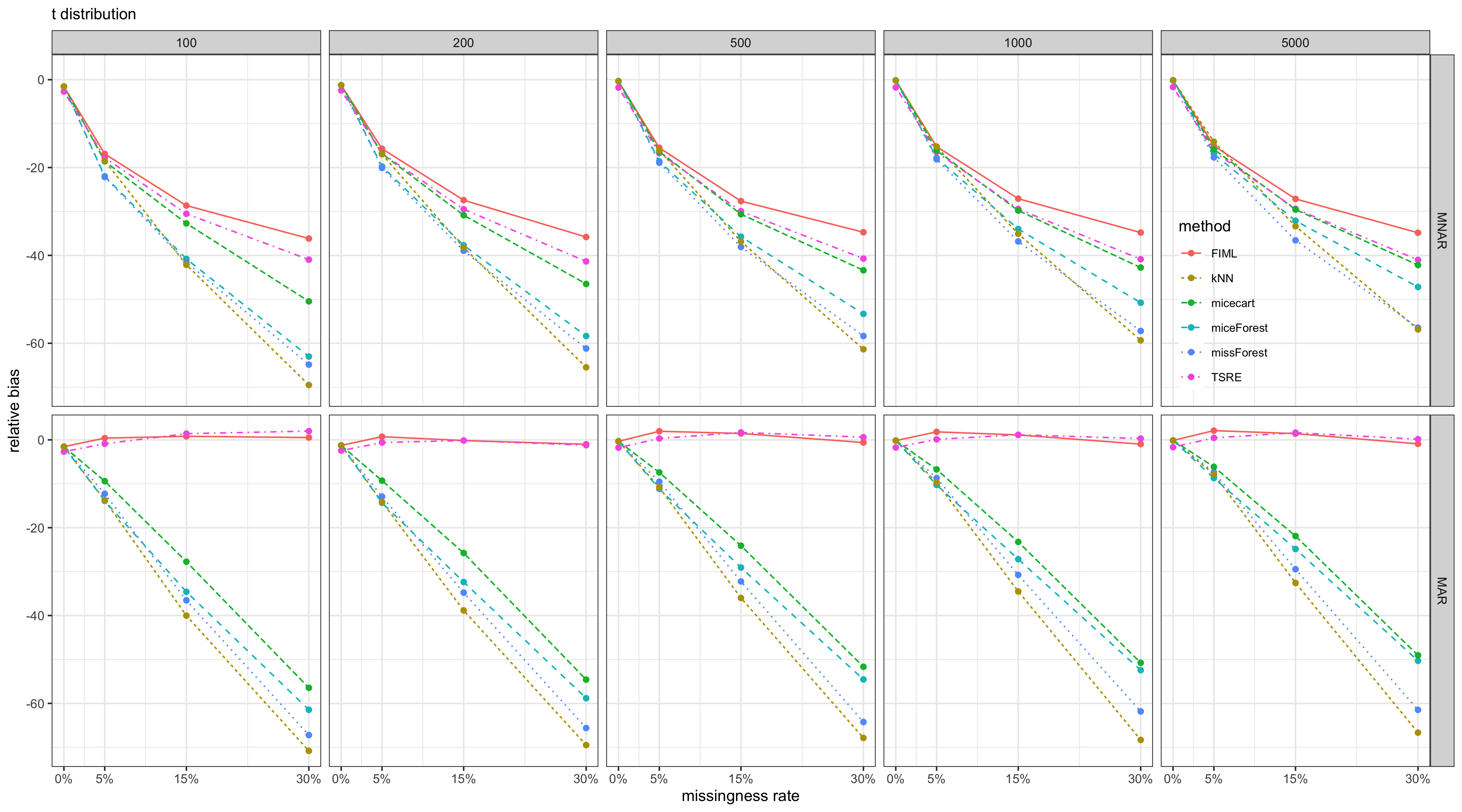} 
    
    \vspace{0.2cm} 
    
    \includegraphics[scale=0.1]{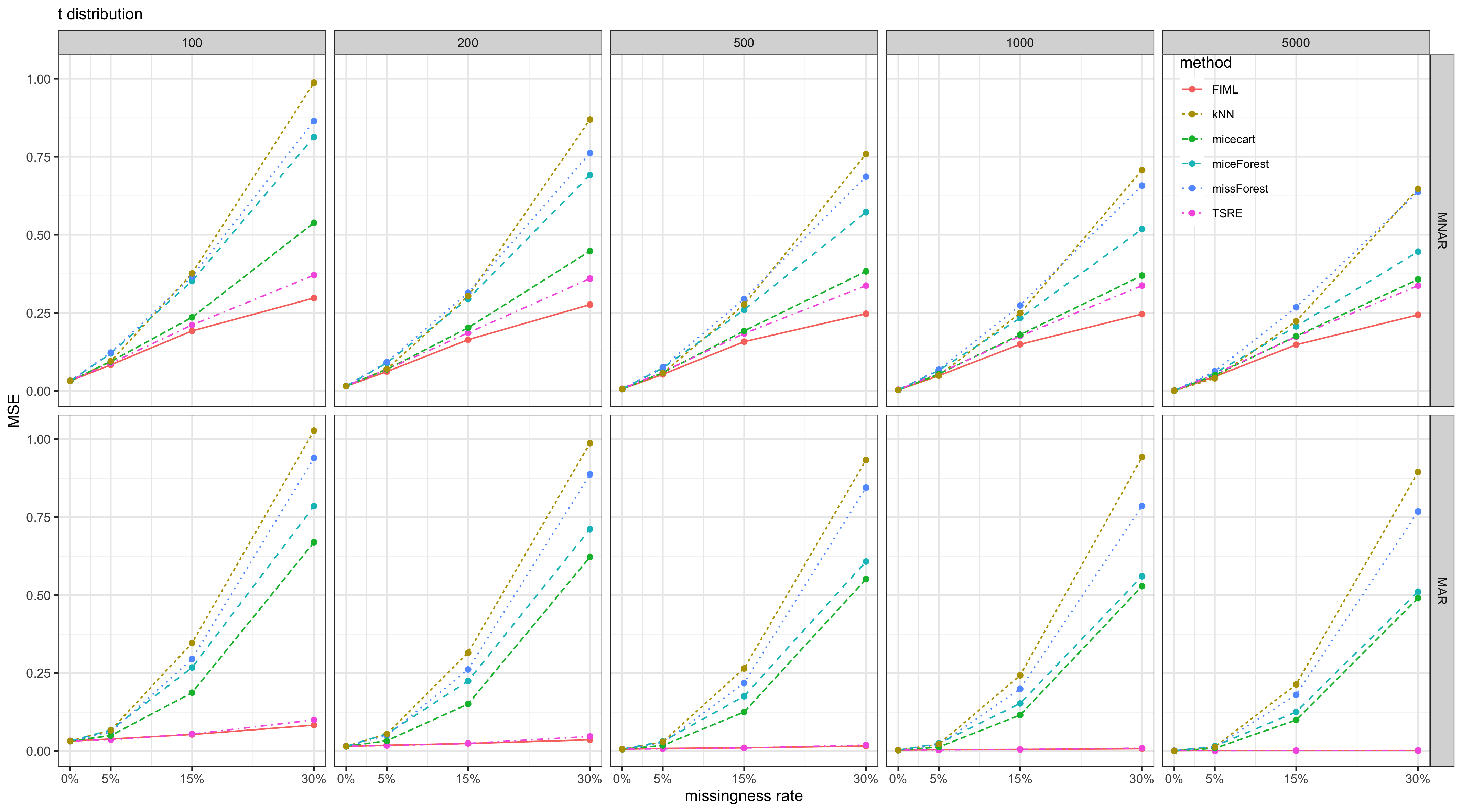} 
    \caption{The relative bias and mean squared error for the average of the variance of the latent slope estimates under the t-distribution} 
    \label{fig:image7}
\end{figure}

\begin{figure}[h]
    \centering
    \includegraphics[scale=0.1]{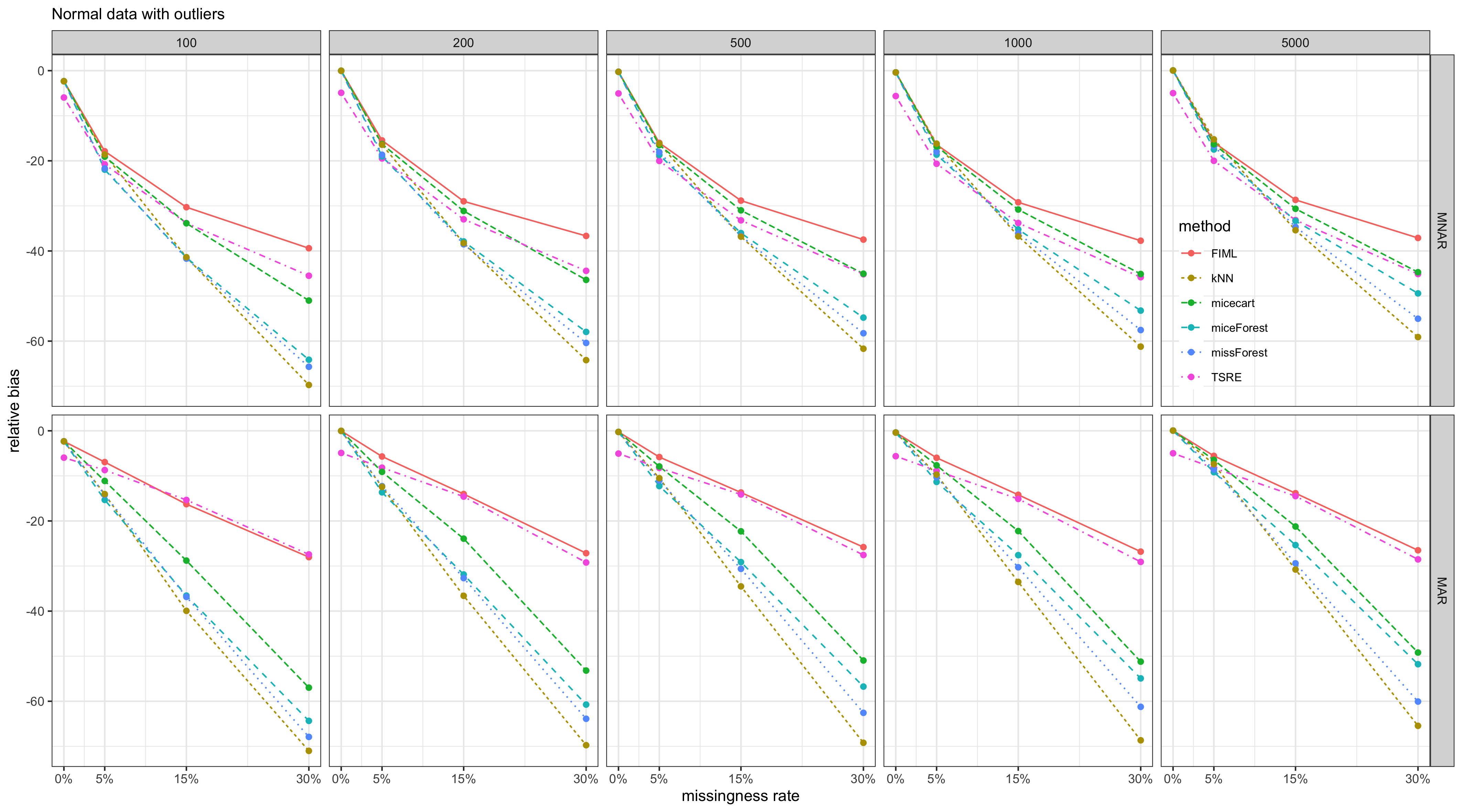} 
    
    \vspace{0.2cm} 
    
    \includegraphics[scale=0.1]{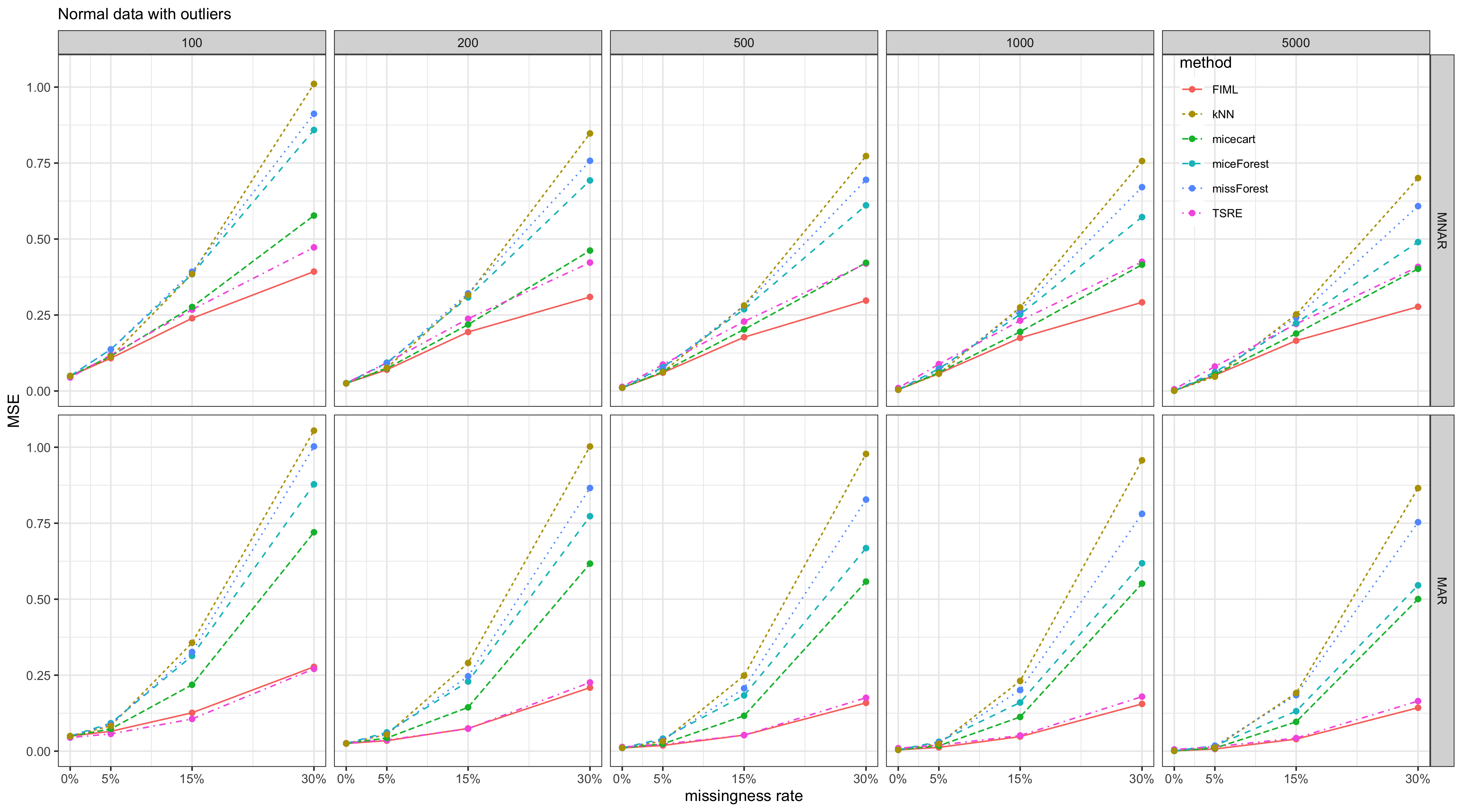} 
    \caption{The relative bias and mean squared error for the average of the variance of the latent slope estimates under the normal distribution with outliers} 
    \label{fig:image8}
\end{figure}
\section{An Empirical Example}
After assessing the performance of six missing data analytical approaches, we illustrate the application of these six approaches in addressing a real world missing data problem in growth curve modeling.  The data are a subset from the National Longitudinal Survey of Youth 1997 Cohort (NLSY97) (Bureau of Labor Statistics, U.S. Department of Labor, 2005), where 399 schoolchildren's Peabody Individual Achievement Test (PIAT) math scores were measured yearly from 1997 to 2000.  

Table 2 shows the descriptive statistics of the math scores at each time point, including mean, skewness, kurtosis, Shapiro–Wilk normality test results, and missingness rate (ranging from 5.514\% to 12.281 \%).  Determining whether the population follows a normal distribution is challenging, primarily because any nonnormality in the data could be attributed to MAR or MNAR missing mechanisms. At the sample level, Shapiro–Wilk normality tests (Shapiro \& Wilk, 1965) indicate that the PIAT math scores from 1997 to 2000 all deviate from normal distributions. Consequently, it is inappropriate to consider the observed data as normally distributed.  
\begin{table}[ht]
	\centering
	\caption{Descriptive statistics and Shapiro-Wilk normality test}
	\label{t2} 
	\begin{tabular}{rrrrrr}
		\hline
		variable & mean & skewness & kurtosis & Shapiro.test & missingness rate\\ 
		\hline
		y1 & 61.16 & -0.08 & 0.21  & 0.99* & 6.02\%\\ 
		y2 & 63.27 & -0.66 & 1.01  &0.97*** & 5.51\%\\ 
		y3 & 67.56 & -0.43  & 0.56 &0.98** & 10.53\%\\ 
        y4  & 69.69  & -0.62 &0.77 &0.96*** & 12.28\%\\  
		\hline
	\end{tabular}
 \begin{tablenotes}
        \footnotesize{*: p \textless .05; **: p \textless .01; ***: p \textless .001.}
      \end{tablenotes}
\end{table}

%\centerline{[ Insert Table 2 here ]}
We used a linear growth curve model to analyze the data and applied six missing data analytical approaches: FIML, TSRE, KNN, micecart, missforest, and miceforest, to handle missing values. The parameter estimates of the growth curve model, as well as their corresponding standard error (SE) estimates using the six missing data analytical approaches, are reported in Table 3. Based on the simulation results, if the missing data mechanism is MAR,  TSRE should be used. If the missing data are MNAR, FIML could still be used when the missingness rate is small. When the missingness rate is large, all six methods cannot be trusted. In practice, it is impossible to statistically test whether the missingness is MAR or MNAR. But fortunately, in this empirical example, the missing data rate is small (less than 13\%).  Our simulation study showed that when the missing data rate is smaller than 15\%, FIML and TSRE have similar performance in parameter estimation. Therefore, we chose TSRE to estimate model parameters, especially because data are potentially nonnormally distributed and skewed.  

Based on TSRE, the average initial mathematical ability in 1997 is approximately 61.04, with a yearly growth rate of 3.20 from 1997 to 2000. There are intraindividual variations in both the initial level and the growth rate. The negative estimates of the covariance between the latent intercept and slope suggest that students with higher initial levels of mathematical ability tend to exhibit lower growth rates, but such relationship is not significant.

Note that the variance of the random slopes is not statistically significant under micecart, indicating that there are no interindividual differences in the growth rate. However, the other five methods all suggest that there are interindividual differences. Using different missing data analytical methods may lead to contradicted conclusions.

%\centerline{[ Insert Table 3 here ]}
\begin{table}[ht]
	\centering
	\caption{Parameter and standard error estimates of the growth curve model using six missing data analytical techniques}
	\label{t3}
    \begin{threeparttable}
	\begin{tabular}{rrrrrr}
		\hline
		& $\beta_L$ (SE) & $\beta_S$ (SE) & $\sigma_L^2$ (SE) & $\sigma_S^2$ (SE) & $\sigma_{LS}$ (SE)\\
		\hline
		\multirow{2}*{FIML} & 60.62*** & 3.10*** & 177.65*** & 7.31* & -5.18 \\
             & (0.79) &  (0.27) & (18.54) & (2.85)& (5.55)\\
		\multirow{2}*{TSRE} & 61.04*** & 3.20*  & 170.84*** & 5.83* & -5.25 \\
             & (0.79) & (0.25) & (19.11)& (2.80)& (4.90)\\
		\multirow{2}*{KNN} & 60.70*** & 3.12*** & 173.99*** & 6.65** & -4.57 \\
            & (0.77) & (0.25) & (17.77)& (2.44) & (5.05)\\
        \multirow{2}*{micecart} & 60.51*** & 3.21*** & 178.54***  & 4.69 & -4.05 \\
                 & (0.78) & (0.25)& (18.67) & (2.62)& (5.37)\\
		\multirow{2}*{missforest} & 60.56*** & 3.17*** & 179.68*** & 5.84* & -5.56 \\
                   & (0.78)& (0.25)& (18.15)& (2.41)& (5.07)\\
		\multirow{2}*{miceforest} & 60.62*** & 3.13*** & 175.69*** & 5.82* & -4.13  \\
                   & (0.78)& (0.26)& (18.45)& (2.64)& (5.36) \\
		\hline
	\end{tabular}
 \begin{tablenotes}
        \footnotesize{$\beta_L$: average latent intercepts;
$\beta_S$: average latent slopes;
$\sigma_L^2$: variance of latent intercept;
$\sigma_S^2$: variance of latent slope;
$\sigma_{LS}$: Covariance between the latent intercept and slope.}
      \end{tablenotes}
     \end{threeparttable}
\end{table}

\section{Discussion}
This study used a Monte Carlo simulation to systematically evaluate traditional and machine learning approaches for handling ignorable and nonignorable missing data in growth curve modeling. Contrary to common perceptions about the superiority of machine learning techniques in this context, our results revealed that, in general, TSRE was proved to be the preferred approach for handling MAR missing data. FIML was the most effective approach for MNAR missing data among the six methods assessed. The missForest technique could be the approach of choice only in specific scenarios: a combination of highly skewed data (lognormal data), very large sample sizes (e.g., $n \geq 1000$), and low missing data rates for the MNAR mechanism. However, a large sample size is rarely seen in longitudinal studies in social and behavioral sciences. This finding suggests that machine learning approaches like KNN, micecart, missForest, and miceForest may not be suitable for missing data analysis in the context of growth curve modeling in practice. 

Although our conclusion regarding the performance of machine learning imputation approaches differs from the literature trend, several existing studies also found that traditional approaches performed better than machine learning approaches under certain conditions. For example, Hong and Lynn (2020) reported that missForest was ineffectual with MAR data with highly skewed distributions. Wu (2023) found that machine learning-based imputation approaches did not outperform traditional parametric imputation in enhancing model performance. Xiao and Bulut (2020) concluded that FIML outperformed micecart and miceForest in handling sparse data with MCAR, MAR, and MNAR missing mechanisms. In addition, miceForest could not adequately handle MAR and MCAR missing data within the structural equation modeling framework (Jia \& Wu, 2019; Jia \& Wu, 2023).

Several factors may explain our findings and justify our conclusion. First of all, most studies investigating machine learning imputation approaches limited their scope to MCAR missing data mechanism (Emmanuel et al., 2021; Stekhoven \& Bühlmann, 2012). So, they may indeed perform well for MCAR data but fail to address nonignorable missingness (Hong \& Lynn, 2020). Second, some existing studies focused on categorical data (Dagdoug et al., 2023; Emmanuel et al., 2021; Pujianto et al., 2019). Although machine learning imputation approaches may adeptly account for categorical data, such techniques are less applicable in psychological research, where most constructs are regarded as continuous (Lin et al., 2019, 2020; Tang \& Wen, 2020). Third, most existing studies exclusively concentrated on cross-sectional data following normal distributions, disregarding longitudinal nonnormal data (Doove et al., 2014; Stekhoven \& Bühlmann, 2012; Tang \& Ishwaran, 2017)—the primary interest of our study. Fourth, psychological and educational modeling often involves latent variables influencing the relationships between observed variables. Machine learning imputation approaches may fail to account for such relationships since latent variables are not included in the analysis. Additionally, in our simulation study, the data generation and the data analytical models were the same, meaning that the model was correctly specified. FIML and TSRE, being model-based approaches, can more accurately capture the underlying relationships in the data and model missing data based on the model. However, the imputation models in the machine learning approaches are different from the data generation model. They thus can not handle missing data as well as the model-based approaches. In reality, the data generation model is unknown. So, the data analytical model could be misspecified. In such cases, machine learning approaches might work relatively better than traditional approaches, but such scenarios need further investigation.

Our study also found that, in general, multiple imputation approaches outperformed the single imputation approaches, which is consistent with most previous studies. But for low missingness rates, large sample sizes and highly skewed data with nonignorable missingness, single imputation approaches, KNN and missForest, could perform better than multiple imputation approaches, such as micecart.  While previous studies indicated that micecart performed better than KNN and missForest (Waljee et al., 2013; Wei et al., 2018), they did not study their performance for MNAR data. Future studies may further explore the performance of KNN and missForest for handling MNAR missing data, particularly with large sample sizes and highly skewed data.

Addressing MNAR missing data remains a significant challenge. Although we showed that FIML performed better than the other five approaches in dealing with MNAR data even when data are not normal, FIML generated acceptably biased estimates only when the missingness rate was low. So, choosing FIML is like choosing the best from a bad lot. This study showed that although FIML led to biased parameter estimates for MNAR data, it still performed better than the machine learning approaches, indicating that researchers should not use these machine learning approaches to impute missing values in longitudinal research. When data are MNAR, more sophisticated approaches, such as selection models or pattern mixture models (Tong et al., 2024), should be adopted to account for the nature of the missingness. 

Note that our simulation study included 140 data conditions and considered four different hyperparameters for the machine learning approaches. For each data condition, we generated 500 replications, and all six missing data analytical techniques were applied to each replication. In total, there were 1,750,000 independent analyses. Due to this large number, we did not manipulate other factors, such as the number of measurement occasions and the covariance between latent intercept and slope. We conducted pilot simulations to examine the impact of the number of measurement occasions on the performance of the machine learning approaches under MNAR. Our findings confirmed that the relative performance of the six missing data analytical methods was not influenced by the number of measurement occasions. These additional simulation results are available on our \href{https://github.com/DandanTang0/Evaluation-of-Missing-Data-Analytical-Techniques-in-Longitudinal-Research}{GitHub} site.

Our simulation study focused on continuous data. In practice, ordinal or categorical data may also be present or may coexist alongside continuous data. Future research should explore whether our findings extend to varied data types, including continuous, ordinal, or a combination of both, to determine if the same conclusions are applicable across different data structures.

\section{References}
\begin{description}
% remove Siciliano, Aria \& D’Ambrosio, 2006   Harrell, 2001;
\item Asparouhov, T., \& Muthén, B. (2014). Multiple-group factor analysis alignment. \emph{Structural Equation Modeling: A Multidisciplinary Journal, 21(4)}, 495-508. https://doi.org/10.1080/10705511.2014.919210 
\item Bárcena, M. J., \& Tusell, F. (2000). Tree-based algorithms for missing data imputation. In \textit{COMPSTAT: Proceedings in Computational Statistics 14th Symposium held in Utrecht, The Netherlands, 2000} (pp. 193-198). Physica-Verlag HD. 
\item Berger, J. O., \& Pericchi, L. R. (2004). Training samples in objective Bayesian model selection. \emph{The Annals of Statistics, 32(3)}, 841-869. https://doi.org/ 10.1214/009053604000000229
\item Breiman, L. (2001). Random forests. \textit{Machine learning}, \textit{45}, 5-32. 
\item Breiman, L., Friedman, J., Olshen, R., \& Stone, C. (1984). Cart. \textit{Classification and regression trees}.
\item Cain, M. K., Zhang, Z., \& Yuan, K. H. (2017). Univariate and multivariate skewness and kurtosis for measuring nonnormality: Prevalence, influence and estimation. \textit{Behavior research methods}, \textit{49}, 1716-1735.
 \item Camerini, A. L., Marciano, L., Carrara, A., \& Schulz, P. J. (2020). Cyberbullying perpetration and victimization among children and adolescents: A systematic review of longitudinal studies. \textit{Telematics and informatics}, \textit{49}, 101362.
\item Conversano, C., \& Siciliano, R. (2009). Incremental tree-based missing data imputation with lexicographic ordering. Journal of classification, 26, 361-379.
 \item Creel, D. V., \& Krotki, K. (2006). Creating imputation classes using classification tree methodology. In Proc. Surv. Res. Methods Sect., Am. Stat. Assoc (pp. 2884-2887).
 \item Cruz, K. L. D., Kelsey, C. M., Tong, X., \& Grossmann, T. (2023). Infant and maternal responses to emotional facial expressions: A longitudinal study. \textit{Infant Behavior and Development}, \textit{71}, 101818.
\item Dagdoug, M., Goga, C., \& Haziza, D. (2023). Imputation procedures in surveys using nonparametric and machine learning methods: an empirical comparison. \textit{Journal of Survey Statistics and Methodology}, \textit{11}(1), 141-188. 
\item Deering, S., Pratap, A., Suver, C., Borelli Jr, A. J., Amdur, A., Headapohl, W., \& Stepnowsky, C. J. (2020). Real-world longitudinal data collected from the SleepHealth mobile app study. \textit{Scientific data}, \textit{7}(1), 418.
\item Doove, L. L., Van Buuren, S., \& Dusseldorp, E. (2014). Recursive partitioning for missing data imputation in the presence of interaction effects. \emph{Computational statistics \& data analysis, 72}, 92-104.
\item Drake, R. J., Husain, N., Marshall, M., Lewis, S. W., Tomenson, B., Chaudhry, I. B., ... \& Birchwood, M. (2020). Effect of delaying treatment of first-episode psychosis on symptoms and social outcomes: a longitudinal analysis and modelling study. \textit{The Lancet Psychiatry}, \textit{7}(7), 602-610.
\item Emmanuel, T., Maupong, T., Mpoeleng, D., Semong, T., Mphago, B., \& Tabona, O. (2021). A survey on missing data in machine learning. \textit{Journal of Big Data}, \textit{8}(1), 1-37.
\item Enders, C. K. (2001). The impact of nonnormality on full information maximum-likelihood estimation for structural equation models with missing data. \emph{Psychological methods, 6(4)}, 352.
\item Enders, C. K. (2010). \emph{Applied missing data analysis}. New York, NY: Guilford.
\item Enders, C. K. (2023). Missing data: An update on the state of the art. \textit{Psychological Methods}. 
\item Enders, C. K., \& Bandalos, D. L. (2001). The relative performance of full information maximum likelihood estimation for missing data in structural equation models. \emph{Structural equation modeling, 8(3)}, 430-457.
\item Finch, W. H., Finch, M. E. H., \& Singh, M. (2016). Data imputation algorithms for mixed variable types in large scale educational assessment: a comparison of random forest, multivariate imputation using chained equations, and MICE with recursive partitioning. International Journal of Quantitative Research in Education, 3(3), 129-153.
\item Fix, E., \& Hodges, J. L. (1952). Discriminatory analysis: Nonparametric discrimination: Small sample performance.
\item Graham, J. W., Olchowski, A. E., \& Gilreath, T. D. (2007). How many imputations are really needed? Some practical clarifications of multiple imputation theory. \textit{Prevention science}, \textit{8}, 206-213.
\item Gower, J. C. (1971). A general coefficient of similarity and some of its properties. Biometrics, 857-871.
 \item Harrell, F. E. (2001). Regression modeling strategies: with applications to linear models, logistic regression, and survival analysis (Vol. 608). New York: springer.
\item Hayes, T., Usami, S., Jacobucci, R., \& McArdle, J. J. (2015). Using Classification and Regression Trees (CART) and random forests to analyze attrition: Results from two simulations. \emph{Psychology and aging, 30(4)}, 911.
\item Ho, T. K. (1995, August). Random decision forests. In Proceedings of 3rd international conference on document analysis and recognition (Vol. 1, pp. 278-282). IEEE.
\item Hong, S., \& Lynn, H. S. (2020). Accuracy of random-forest-based imputation of missing data in the presence of non-normality, non-linearity, and interaction. \textit{BMC medical research methodology}, \textit{20}(1), 1-12.
\item Hoogland, J. J., \& Boomsma, A. (1998). Robustness studies in covariance structure modeling: An overview and a meta-analysis. \textit{Sociological Methods \& Research}, \textit{26}(3), 329-367.
\item Huang, L., Abrahams, A., \& Ractham, P. (2022). Enhanced financial fraud detection using cost‐sensitive cascade forest with missing value imputation. \textit{Intelligent Systems in Accounting, Finance and Management}, \textit{29}(3), 133-155.
\item Javadi, S., Bahrampour, A., Saber, M. M., Garrusi, B., \& Baneshi, M. R. (2021). Evaluation of four multiple imputation methods for handling missing binary outcome data in the presence of an interaction between a dummy and a continuous variable. \textit{Journal of Probability and Statistics}, \textit{2021}, 1-14.
\item Jerez, J. M., Molina, I., García-Laencina, P. J., Alba, E., Ribelles, N., Martín, M., \& Franco, L. (2010). Missing data imputation using statistical and machine learning methods in a real breast cancer problem. \textit{Artificial intelligence in medicine}, \textit{50}(2), 105-115. 
\item Jia, F., \& Wu, W. (2019). Evaluating methods for handling missing ordinal data in structural equation modeling. \textit{Behavior Research Methods}, \textit{51}, 2337-2355.
\item Jia, F., \& Wu, W. (2023). A comparison of multiple imputation strategies to deal with missing nonnormal data in structural equation modeling. \textit{Behavior Research Methods}, \textit{55}(6), 3100-3119.
\item Jing, X., Luo, J., Wang, J., Zuo, G., \& Wei, N. (2022). A Multi-imputation method to deal with hydro-meteorological missing values by integrating chain equations and random forest. \textit{Water Resources Management}, \textit{36}(4), 1159-1173.
 \item Jonsson, P., \& Wohlin, C. (2004, September). An evaluation of k-nearest neighbour imputation using likert data. In \textit{10th International Symposium on Software Metrics, 2004. Proceedings.} (pp. 108-118). IEEE.
\item Kowarik, A., \& Templ, M. (2016). Imputation with the R Package VIM. \textit{Journal of statistical software}, \textit{74}, 1-16. 
\item Ibrahim, J. G., \& Molenberghs, G. (2009). Missing data methods in longitudinal studies: a review. \textit{Test}, \textit{18}(1), 1-43.
\item Itzchakov, G., Weinstein, N., Vinokur, E., \& Yomtovian, A. (2023). Communicating for workplace connection: A longitudinal study of the outcomes of listening training on teachers' autonomy, psychological safety, and relational climate. \textit{Psychology in the Schools}, \textit{60}(4), 1279-1298.
\item Jönsson, P., \& Wohlin, C. (2006). Benchmarking k-nearest neighbour imputation with homogeneous Likert data. \textit{Empirical Software Engineering}, \textit{11}, 463-489. 
\item Laqueur, H. S., Shev, A. B., \& Kagawa, R. M. (2022). SuperMICE: An ensemble machine learning approach to multiple imputation by chained equations. \textit{American journal of epidemiology}, \textit{191}(3), 516-525.
\item Leeper, J. D., \& Woolson, R. F. (1982). Testing hypotheses for the growth curve model when the data are incomplete. \textit{Journal of Statistical Computation and Simulation}, \textit{15}(2-3), 97-107. 
\item Lei, P. W., \& Shiverdecker, L. K. (2020). Performance of estimators for confirmatory factor analysis of ordinal variables with missing data. \emph{Structural Equation Modeling: A Multidisciplinary Journal, 27(4)}, 584-601.
\item Lin, X. F., Tang, D., Lin, X., Liang, Z. M., \& Tsai, C. C. (2019). An exploration of primary school students’ perceived learning practices and associated self-efficacies regarding mobile-assisted seamless science learning. \emph{International Journal of Science Education, 41(18)}, 2675-2695.
\item Lin, X. F., Tang, D., Shen, W., Liang, Z. M., Tang, Y., \& Tsai, C. C. (2020). Exploring the relationship between perceived technology-assisted teacher support and technology-embedded scientific inquiry: the mediation effect of hardiness. \emph{International Journal of Science Education, 42(8)}, 1225-1252.
\item Little, R. J. A., \& Rubin, D. B. (2002). \emph{Statistical analysis with missing data} (2nd ed.). New York, NY: Wiley-Interscience
\item  Loh, W. Y. (2011). Classification and regression trees. \textit{Wiley interdisciplinary reviews: data mining and knowledge discovery}, \textit{1}(1), 14-23. 
\item Micceri, T. (1989). The unicorn, the normal curve, and other improbable creatures. \textit{Psychological bulletin}, \textit{105}(1), 156. 
\item Muthén, L. K., \& Muthén, B. (2017). \emph{Mplus user's guide: Statistical analysis with latent variables}
\item Parr, C. L., Hjartåker, A., Scheel, I., Lund, E., Laake, P., \& Veierød, M. B. (2008). Comparing methods for handling missing values in food-frequency questionnaires and proposing k nearest neighbours imputation: effects on dietary intake in the Norwegian Women and Cancer study (NOWAC). Public health nutrition, 11(4), 361-370.
 \item Pujianto, U., Wibawa, A. P., \& Akbar, M. I. (2019, October). K-nearest neighbor (k-NN) based missing data imputation. In \textit{2019 5th International Conference on Science in Information Technology (ICSITech)} (pp. 83-88). IEEE.
\item Raykov, T. (2005). Analysis of longitudinal studies with missing data using covariance structure modeling with full-information maximum likelihood. \emph{Structural Equation Modeling, 12}, 493–505.
\item Ribeiro, C., \& Freitas, A. A. (2019). Comparing the effectiveness of six missing value imputation methods for longitudinal classification datasets. In \textit{3rd Workshop on AI for Aging, Rehabilitation and Independent Assisted Living (ARIAL), held as part of IJCAI-2019}.
\item Rizvi, S. T. H., Latif, M. Y., Amin, M. S., Telmoudi, A. J., \& Shah, N. A. (2023). Analysis of Machine Learning Based Imputation of Missing Data. \textit{Cybernetics and Systems}, 1-15.
\item Rodriguez, D. (2023). Area under the Curve as an Alternative to Latent Growth Curve Modeling When Assessing the Effects of Predictor Variables on Repeated Measures of a Continuous Dependent Variable. \emph{Stats, 6(2)}, 674-688.
\item Rosseel, Y. (2012). \emph{lavaan: a brief user’s guide}. 
\item Rubin, D. B. (1976). Inference and missing data. \emph{Biometrika, 63}, 581–592.
 \item Sania, A., Pini, N., Nelson, M., Myers, M. M., Shuffrey, L. C., Lucchini, M., ... \& Fifer, W. (2021). The K nearest neighbor algorithm for imputation of missing longitudinal prenatal alcohol data. \textit{Available at SSRN 4065215}.
\item Savalei, V., \& Bentler, P. M. (2009). A two-stage approach to missing data: Theory and application to auxiliary variables. \emph{Structural Equation Modeling, 16}, 477–497.
\item Serang, S., \& Sears, J. (2021). Tree-based Matching on Structural Equation Model Parameters. \emph{Journal of Behavioral Data Science, 1(2)}, 31-53. 
\item  Schaffer, C. (1993). Overfitting avoidance as bias. \textit{Machine learning}, \textit{10}, 153-178. 
\item Schafer, J. L., \& Graham, J. W. (2002). Missing data: our view of the state of the art. \textit{Psychological methods}, \textit{7}(2), 147. 
\item Shah, A. D., Bartlett, J. W., Carpenter, J., Nicholas, O., \& Hemingway, H. (2014). Comparison of random forest and parametric imputation models for imputing missing data using MICE: a CALIBER study. \textit{American journal of epidemiology}, \textit{179}(6), 764-774.
\item Shin, T., Davison, M. L., \& Long, J. D. (2017). Maximum likelihood versus multiple imputation for missing data in small longitudinal samples with nonnormality. \textit{Psychological methods}, \textit{22}(3), 426. 
\item Siciliano, R., Aria, M., \& D’Ambrosio, A. (2006). Boosted incremental tree-based imputation of missing data. In Data Analysis, Classification and the Forward Search: Proceedings of the Meeting of the Classification and Data Analysis Group (CLADAG) of the Italian Statistical Society, University of Parma, June 6–8, 2005 (pp. 271-278). Springer Berlin Heidelberg.
\item Silva, K., \& Miller, V. A. (2022). Parenting and the development of impulse control in youth with type 1 diabetes: the mediating role of negative affect. \textit{Applied developmental science}, \textit{26}(1), 94-108. 
\item Smith, B. I., Chimedza, C., \& Bührmann, J. H. (2021). Random forest missing data imputation methods: Implications for predicting at-risk students. In Intelligent Systems Design and Applications: 19th International Conference on Intelligent Systems Design and Applications (ISDA 2019) held December 3-5, 2019 19 (pp. 298-308). Springer International Publishing.
\item Stegmann, G., Jacobucci, R., Serang, S., \& Grimm, K. J. (2018). Recursive partitioning with nonlinear models of change. \emph{Multivariate behavioral research, 53(4)}, 559-570.
\item Stekhoven, D. J., \& Bühlmann, P. (2012). MissForest—non-parametric missing value imputation for mixed-type data. \emph{Bioinformatics, 28(1)}, 112-118.
\item Tang, D., Boker, S. M., \& Tong, X. (2024). Are the Signs of Factor Loadings Arbitrary in Confirmatory Factor Analysis? Problems and Solutions. \emph{Structural Equation Modeling: A Multidisciplinary Journal}, 1-13.
 \item Tang, D., \& Tong, X. (2023). A Comparison of Full Information Maximum Likelihood and Machine Learning Missing Data Analytical Methods in Growth Curve Modeling. \textit{arXiv preprint arXiv:2312.17363}.
 \item Tang, D. \& Wen, Z. (2020). Statistical approaches for testing common method bias: Problems and suggestions. \emph{Journal of Psychological Science, 43(1)}, 215-223.
\item Tang, F., \& Ishwaran, H. (2017). Random forest missing data algorithms. \textit{Statistical Analysis and Data Mining: The ASA Data Science Journal}, \textit{10}(6), 363-377. 
\item Therneau, T., Atkinson, B., Ripley, B., \& Ripley, M. B. (2015). Package ‘rpart’. https://cran.r-project.org/web/packages/rpart/rpart.pdf.
 \item Tong, X., Kim, S., Bandyopadhyay, D., \& Sun, S. (2024). Association Between Body Fat and Body Mass Index from Incomplete Longitudinal Proportion Data: Findings from the Fels Study. \textit{Journal of Data Science}, \textit{22}(1).
\item Tong, X., Zhang, Z., \& Yuan, K. H. (2014). Evaluation of test statistics for robust structural equation modeling with nonnormal missing data. \emph{Structural Equation Modeling: A Multidisciplinary Journal, 21(4)}, 553-565.
\item Tong, X., Zhang, T., \& Zhou, J. (2021). Robust Bayesian growth curve modeling using conditional medians. \emph{British Journal of Mathematical and Statistical Psychology, 74(2)}, 286-312.
\item van Buuren, S. (2012). Flexible Imputation of Missing Data.
\item Van Buuren, S., \& Groothuis-Oudshoorn, K. (2011). mice: Multivariate imputation by chained equations in R. Journal of statistical software, 45, 1-67.
 % \item Wallace, M. L., Anderson, S. J., \& Mazumdar, S. (2010). A stochastic multiple imputation algorithm for missing covariate data in tree‐structured survival analysis. Statistics in Medicine, 29(29), 3004-3016.
 \item Wang, Z., Akande, O., Poulos, J., \& Li, F. (2022). Are deep learning models superior for missing data imputation in surveys? Evidence from an empirical comparison. \textit{Survey Methodology}, \textit{48}(2), 375-399.
 \item Waljee, A. K., Mukherjee, A., Singal, A. G., Zhang, Y., Warren, J., Balis, U., ... \& Higgins, P. D. (2013). Comparison of imputation methods for missing laboratory data in medicine. \textit{BMJ open}, \textit{3}(8), e002847. 
\item Wei, R., Wang, J., Su, M., Jia, E., Chen, S., Chen, T., \& Ni, Y. (2018). Missing value imputation approach for mass spectrometry-based metabolomics data. \textit{Scientific reports}, \textit{8}(1), 663. 
\item Wen, Z., Huang, B., \& Tang, D. (2018). Preliminary work for modeling questionnaire data. \emph{Journal of Psychological Science, 41(1)}, 204-210.
\item Wen, Z., Tang, D., \& Gu, H. (2019). A general simulation comparison of the predictive validity between bifactor and high-order factor models. \emph{Acta Psychologica Sinica, 51(3)}, 383-391.
\item Wongkamthong, C., \& Akande, O. (2023). A comparative study of imputation methods for multivariate ordinal data. \textit{Journal of Survey Statistics and Methodology}, \textit{11}(1), 189-212.
 \item Wu, Y. (2023). \textit{Handling missing values in risk prediction modeling: a comparative simulation study on parametric and machine learning multiple imputations} (Doctoral dissertation).
\item Xiao, J., \& Bulut, O. (2020). Evaluating the performances of missing data handling methods in ability estimation from sparse data. Educational and Psychological Measurement, 80(5), 932-954.
\item Yuan, K. H. (2009). Normal distribution based pseudo ML for missing data: With applications to mean and covariance structure analysis. \emph{Journal of Multivariate Analysis, 100(9)}, 1900-1918.
\item Yuan, K. H., \& Bentler, P. M. (2000). 5. Three likelihood-based methods for mean and covariance structure analysis with nonnormal missing data. \emph{Sociological methodology, 30(1)}, 165-200.
\item Yuan, K. H., \& Bentler, P. M. (2001). Effect of outliers on estimators and tests in covariance structure analysis. \emph{British Journal of Mathematical and Statistical Psychology, 54(1)}, 161-175.
\item Yuan, K. H., Tong, X., \& Zhang, Z. (2015). Bias and efficiency for SEM with missing data and auxiliary variables: Two-stage robust method versus two-stage ML. Structural Equation Modeling: A Multidisciplinary Journal, 22(2), 178-192.
\item Yuan, K. H., \& Zhang, Z. (2012). Robust structural equation modeling with missing data and auxiliary variables. \emph{Psychometrika, 77(4)}, 803-826.
\item Zhang, S. (2012). Nearest neighbor selection for iteratively kNN imputation. \textit{Journal of Systems and Software}, \textit{85}(11), 2541-2552.   %这个研究写了single and multiple imputations
 \item Zhang, S., Gong, L., Zeng, Q., Li, W., Xiao, F., \& Lei, J. (2021). Imputation of gps coordinate time series using missforest. \textit{Remote Sensing}, \textit{13}(12), 2312.
\item Zhu, X., Yang, Y., Xiao, Z., Pooley, A. L., Ozdemir, E., Speyer, L. G., ... \& Murray, A. L. (2023). Daily life emotion regulation processes as transdiagnostic predictors of mental health symptoms: An ecological momentary assessment study.
\end{description} 

\end{document}